\documentclass[5p,times]{elsarticle}

\usepackage[final]{changes}
\definechangesauthor{alex}
\definechangesauthor{anna}
\definechangesauthor{toni}

\usepackage{subfigure}
\usepackage{hyperref}
\usepackage{hyphenat}

\usepackage{listings}

\usepackage{mathtools}
\usepackage{algorithm}
\usepackage{algpseudocode}

\usepackage{siunitx}
\usepackage{todonotes}

\usepackage{tablefootnote}
\usepackage{multirow}

\newcommand{\boilerplatemethod}{ The DRAM execution is done with no assistance of other memory tiers. The NVM execution is done in-place on the persistent memory space. The MM executions correspond to the transparent \emph{Memory Mode} available in the OptaneDC devices, which resembles a caching --hence the hot/cold executions. }
\newcommand{\boilerplateappsmall}{ The dataClay (dC) executions use the active object store in different configurations, while DAOS is the non-active one. The DRAM execution uses no additional memory tier. The NVM execution is done in-place on the persistent memory space. The MM execution uses the transparent mode of the OptaneDC called \emph{Memory Mode.} }
\newcommand{\boilerplateappbig}{ The dataClay (dC) executions use the active object store in different configurations, while DAOS is the non-active one. The big dataset does not fit in DRAM. The NVM execution is done in-place on the persistent memory space, while the MM uses the \emph{Memory Mode} transparent behaviour available in the OptaneDC devices. }

\title{Revisiting Active Object Stores: Bringing Data Locality to the Limit With NVM}

\begin{document}

\author{
  Alex Barcelo\\
  \emph{Barcelona Supercomputing Center}
}
\address{alex.barcelo@bsc.es}

\author{
  Anna Queralt\\
%  Serra H\'unter Fellow\\ % already in acknowledgements
  \emph{Universitat Polit\`ecnica de Catalunya\\
  Barcelona Supercomputing Center}
}
\address{anna.queralt@upc.edu}

\author{
  Toni Cortes\\
  \emph{Universitat Polit\`ecnica de Catalunya\\
  Barcelona Supercomputing Center}
}
\address{toni.cortes@ac.upc.edu}

\begin{abstract}

Object stores are widely used software stacks that achieve excellent scale-out with a well-defined interface and robust performance. However, their traditional \emph{get}/\emph{put} interface is unable to exploit data locality at its fullest, and limits reaching its peak performance. In particular, there is one way to improve data locality that has not yet achieved mainstream adoption: the \emph{active} object store. Although there are some projects that have implemented the main idea of the \emph{active} object store such as Swift's Storlets or Ceph Object Classes, the scope of these implementations is limited.

We believe that there is a huge potential for active object stores in the current \emph{status quo}. Hyper\hyp{}converged nodes are bringing more computing capabilities to storage nodes --and viceversa. The proliferation of non-volatile memory (NVM) technology is blurring the line between system memory (fast and scarce) and block devices (slow and abundant). More and more applications need to manage a sheer amount of data (data analytics, Big Data, Machine Learning \& AI, etc.), demanding bigger clusters and more complex computations. All these elements are potential game changers that need to be evaluated in the scope of \emph{active} object stores.

More specifically, having NVM devices presents additional opportunities, such as in-place execution. Being able to use the NVM from within the storage system while taking advantage of in-place execution (thanks to the byte-addressable nature of the NVM), in conjunction with the computing capabilities of hyper\hyp{}converged nodes, can lead to \emph{active} object stores that greatly outperform their non-active counterparts.

In this article we propose an \emph{active} object store software stack and evaluate it on an NVM-populated node. We will show how this setup is able to reduce execution times from 10\% up to more than 90\% in a variety of representative application scenarios. Our discussion will focus on the \emph{active} aspect of the system as well as on the implications of the memory configuration.

\end{abstract}

\maketitle

\section{Introduction}

\added{
The object store software stack design has proved to be scalable and robust, and has become a widely used technology. Different implementations have proven to be invaluable agents in the explosive growth experienced in the cloud arena.} Furthermore, they are also becoming a key technology for a whole lot of different scenarios, such as HPC~\cite{daos-web,ceph-web}, caching middleware~\cite{memcached}, or data analytics~\cite{rupprecht2017swiftanalytics}.

The canonical way of using an object store is through a \emph{get}/\emph{put} interface. Objects may have some additional schema such as columns or data types~\cite{website_mongodb}, and access to data may include some more fine-grained policies in the form of ACL~\cite{openstack-swift}. Other object stores may define the object as a binary blob with flat permissions~\cite{website_minio}. In all cases, access is done from the application to the object store and it implies a certain data transfer --which may be intra\hyp{}node or inter\hyp{}node, depending on the infrastructure.

This usage procedure is unable to exploit data locality at its fullest. At best, the application will be copying an object that resides in memory --e.g., with an in\hyp{}memory node\hyp{}local object store such as Redis~\cite{website_redis}; at worst, the application will be transferring big data structures through the network and waiting for a high-latency memory tier such as a spinning hard drive. 

In order to avoid those indirections, there is one way to improve data locality that has not yet achieved mainstream adoption: the \emph{active} object store. Although object stores have experienced an upwards popularity in recent years, their \emph{active capabilities} have been only timidly appearing, both in research and in production.

\added{
The \emph{active} feature of an object store enables the execution of code within the object store. For instance, an application could ship the code to evaluate the average of a dataset and let the object store compute it without moving any data from it. Later on Section~\ref{sec:objectstore} we elaborate further on the object store stack and the active capabilities, as well as the synergies between those and the NVM. In the broadest sense, we can classify as \emph{active} object store all object stores that offer certain additional computing capabilities.
}

These active capabilities can vary and have different goals. For instance, the ability to execute management routines --defined for or focused on administration-- is an example of a certain kind of active feature. Active capabilities can also be user\hyp{}defined; an example of this would be Storlets~\cite{openstack-storlets}, an extension for Swift~\cite{openstack-swift}. That design is focused on invocations of user-defined code on \emph{put}, \emph{get}, \emph{copy} operations in a per\hyp{}object basis. For instance, calculating the hash of an object and storing its value as an attribute of the same object, or masking some field for privacy reasons and storing the new value on it.
Although that approach is much more flexible than the mechanisms present in other systems, it still lacks a basic flow to enable application\hyp{}driven calls to routines, arbitrary update of objects, or access to multiple objects from a single execution.

One possible root cause for the scarcity of active storage systems is the long-existing chasm between system memory (byte addressable, volatile, fast, scarce) and storage (composed of persistent but slow block devices, with huge latencies and a smaller bandwidth). This gap is now being blurred thanks to Non-Volatile Memory devices (NVM). Those kind of devices are byte-addressable and have a performance that typically sits between DRAM and SSD; NVM are faster than block devices although that comes with a higher cost-per-byte. They are cheaper than DRAM, and thus less scarce, but they are not as fast. An application can use NVM as a ``fast storage device''; the application can also go one step further and perform in-place execution --thanks to the byte-addressable nature of the NVM. Being able to execute in-place, without data copies to main memory, can have a great impact on data locality and thus performance.

Another issue that may have prevented the adoption of \emph{active} storage systems is the major challenges that they present: resource management becomes much more complex and the storage system must manage an imbricated competition for resources --both storage and computation.

Bringing closer computation and storage through an \emph{active object store} has tremendous performance potential, but those designs should keep in check the interference. The interference endured by storage systems has already been object of study~\cite{yildiz2016root}, and its impact is amplified under an \emph{active} object store, as the compute requirements are increased.
For example, in a multi-user or multi-tenant scenario, one heavy computation execution may impact other time-critical applications, resulting in an undesired resource competition amongst executions.

There is a family of storage systems that tackle the interference issues and some of the aforementioned design challenges with a distinct approach: ephemeral storage systems, such as ad-hoc filesystems~\cite{brinkmann2020adhocfs}. 
\added{Ephemeral systems are designed to be executed next to the computation, as opposed to the more traditional approach of independent storage nodes (shared by multiple applications). In the context of HPC and task-queued systems, this means that those storage systems can be deployed using the same nodes than the job, thus having exclusive access to them avoiding any interference with other jobs. Ephemeral systems can also be used in other environments; the orchestration will differ in a case to case scenario, but in all cases the storage system is deployed on top of the same resources as the application. Because they are more tightly coupled to the application, they can also be optimized for the specific application requirements, and can be implemented and used from user space.}

%Such kind of storage systems can be deployed for a limited lifetime (as small as the runtime for a a single job), they can be optimized for the specific application requirements, and can be implemented and used from user space. The resources used by those ephemeral filesystems are the ones available for the application, avoiding any unnecessary (and potentially harmful) competition for resources.
We can see how this frame changes certain challenges and previously mentioned considerations.
For instance, multi-tenancy and interference is irrelevant for user\hyp{}deployed node local infrastructures; and resource management has small repercussions in the absence of multi\hyp{}application interactions. Those kind of storage systems are able to achieve great performance under the proper conditions.
It follows that an \emph{active ephemeral} object store would ensure an extremely high performance, given its high data locality and its ephemeral strengths. A non-ephemeral setup with similar characteristics is achievable, but it will be more sensitive to interference and will also require efforts to address inherent challenges such as multi-tenancy, resource management, etc.

With all those ideas in mind, this article will propose and discuss an active object store architecture with flexible (user\hyp{}defined and application\hyp{}specific) active routines.
In addition to the fundamental design of the active storage system, we will also study the impact and potential related to the usage of an NVM technology within this environment. This kind of memories are able to leverage data locality thanks to their byte\hyp{}addressable nature. It follows that they can be a valuable asset in our context and greatly improve the performance of data processing applications.

In order to evaluate and demonstrate the potential of the proposed active object store we will use a series of applications and run them in an NVM\hyp{}equipped hyper-converged node, under various memory configurations. 

The main contributions of this paper are the following:

\begin{itemize}
    \item {\bf Revisiting the active object store concept}.
    We will discuss its strengths and its potential to exploit data locality beyond the current state of the art. This study is of special relevance, given the appearance of NVM devices that will change the overhead weights due to their fast access time and their byte addressable nature, enabling in-site computation.
    \item {\bf Evaluation of an active object store integrating NVM technology.}
    We will showcase the characteristics and performance of an actual implementation of an active object store, dataClay~\cite{marti2017dataclay}, leveraging NVM devices to exploit data locality. Our evaluation will use a set of well-known \emph{kernel} applications that serve as representatives of a wide range of applications and use cases.
    \item {\bf The characterization of the performance across different memory configurations.}
    The evaluation will be done with Intel Optane DC, an NVM device supporting different configuration modes. We will evaluate the general byte-addressable access to such memories --in Optane terminology, this is called \emph{App Direct} mode-- as well as an Optane DC exclusive configuration, \emph{Memory Mode}, which uses NVM as system memory and DRAM as a cache layer.
\end{itemize}

%%%%%%%%%%%%%%%%%%% Paper outline
The paper is structured as follows: In Section~\ref{sec:relatedwork} we will review related work and state of the art context of \emph{active} object stores and related storage systems. Section~\ref{sec:objectstore} discusses the notion of the object store, including the \emph{active} capabilities that we will be using as well as the impact and usage strategies of the NVM.
After that, Section~\ref{sec:apps} contains an overview of the different applications that we will be using for the evaluation. The software and hardware configuration will be detailed in Section~\ref{sec:methodology}, just before the evaluation (Section~\ref{sec:evaluation}). Finally, Section~\ref{sec:conclusions} contains the discussion and conclusions reached after analyzing the performance and results obtained.
%%%%%%%%%%%%%%%%%%%

\section{Related work}
\label{sec:relatedwork}

\subsection{Active object stores}
\label{relatedwork:objectstores}

The popularity of object stores has soared in the last decades. They have become ubiquitous in the cloud arena as a flexible solution to the scale-out requirements of the cloud growth. The OpenStack group has Swift~\cite{openstack-swift} as their in-house object store. S3~\cite{amazon-s3}, from the Amazon ecosystem, has become a \emph{de facto} object store standard. Object stores have also entered the HPC ecosystem: Ceph~\cite{weil2006ceph} is a distributed filesystem targeted for high-performance with a built-in object store interface. Intel has also contributed to this field by publishing DAOS, an object store that has been ``designed from the ground up for massively distributed Non Volatile Memory''~\cite{daos-web}.

Extending some of the aforementioned object stores with certain active features has been discussed and subject of experimentation. 
Ceph has the concept of \emph{object classes}~\cite{fisk2017mastering}, a feature that can be used during build/deploy stage of the storage system. These shared object classes are able to improve data locality by the execution of routines (previously deployed into the storage nodes, aka as OSDs) within the storage infrastructure itself instead of the client --which is an \emph{active} feature of the storage system.

Another project that extends an object store with active features is Storlets~\cite{openstack-storlets}, which belongs to OpenStack. Its purpose is to ``extend Swift with the ability to run user defined computations --called storlets-- near the data (...)''. This describes the \emph{active} feature of an object store as a means to improve data locality. This feature can be used during runtime, but their trigger mechanism is limited to specific operations on the object store (get, put, copy) and the executed code can only involve the object that is being accessed.

The aforementioned active features showcase the strengths of an active object store, but they cannot be used for arbitrary application flows. For the Storlets project, the scope is limited due to the trigger mechanisms (not at-will, but limited to the \emph{get}, \emph{put}, \emph{copy} operations on the object store). For the Ceph Object Classes, the deployment of classes has dependencies on the internal functionality of Ceph, which restricts users to build object classes within the tree. These limitations hinder the versatility of those solutions.

In the context of DAOS, the work of Lofstead \emph{et al.}~\cite{lofstead2016daos} proposes an architecture for a future exascale storage system, using DAOS as backend. It is relevant to mention the \emph{Function and Analysis Shipping} mechanisms described in that article. The main idea behind those code shipping mechanisms is the avoidance of data transfers --which results in an improvement on data locality. However, those mechanisms are not integrated into the object store: DAOS is used for persistence (in I/O nodes) while execution is done in the compute nodes on top of HDF5. The goal of improving data locality through the shipping of code is shared with our objectives, but our design goes one step beyond and integrates this code shipping and code execution into the object store, resulting in an \emph{active} object store.

\added{
We can see other storage systems that strive to improve performance by shipping code where data is. The ActiveSpaces framework~\cite{docan2011moving} is an example: a storage system capable of moving (and executing) code where the data is being staged. Warren \emph{et al.}~\cite{warren2019analysis} discuss an object/array centric approach that is able to use shared libraries or application executables in order to perform what we will be calling \emph{active} operations. Those approaches show the advantages of active strategies; however they provide a low-level interface and expect extensive knowledge (from the application developer) regarding the architecture in terms of network and computation resources.
}

\subsection{Ephemeral systems}

\added{
Given the computational requirements that an active object store will have, it is natural to look into the ephemeral designs. Interference is a problem that arises when storage systems are shared and the computation cost increases. By using an ephemeral system, the storage stack is deployed close and exclusive to the application (in the same nodes).
}

A relevant discussion and in-depth analysis on ephemeral systems has already been done by Brinkmann \emph{et al.}~\cite{brinkmann2020adhocfs}; that article discusses the general ideas of ad-hoc file systems as well as the specific characteristics of three implementations: BeeOND~\cite{herold2014introduction}, GekkoFS~\cite{vef2020gekkofs}, and BurstFS~\cite{wang2016ephemeral}. 

\added{
By design, ephemeral systems resources are the ones available for the application; this makes them a perfect candidate for active storage systems, which are specially sensible to interference and resource conflicts. An active storage system does not \emph{have} to be an ephemeral system, but deploying it as such is straightforward and eases certain deployment challenges.
}

\subsection{Leveraging different memory tiers}

The speed of block storage has been an important challenge for storage systems. Compared to system memory, block devices have slow access due to their lower bandwidth and higher latency. That problem has been tackled with different strategies. 
An elementary solution to this problem is to not use the disk at all, which is the basis of the RAM\-Clouds infrastructure~\cite{ousterhout2010case}. 
This is a high performance storage system that achieves great scalability and very low latency by only using the DRAM.

\added{
Data Elevator~\cite{dong2016data} is an example of a software library that attempts to address the problem of the storage hierarchy. The library performs the movement of data between the different memory tiers, and it does so in an efficient way that is transparent from the application point of view.
}

The appearance of non-volatile memory devices created new research lines and also rejuvenated a whole lot of existing ones. For example, Fan \emph{et al.} propose the \emph{Hibachi cache}~\cite{fan2017hibachi} and discuss the improvements of a cooperative cache (which uses DRAM and NVRAM) between different memory tiers.
Hermes~\cite{kougkas2018hermes} is an example of a buffering system which is aware (and takes advantage) of all the different tiers in the storage hierarchy --including NVRAM. In the distributed file systems field there have also been some efforts on improving distributed file systems thanks to the new NVM tier, a topic that has been explored and evaluated by Herodotou and Kakoulli~\cite{herodotou2019automating}. 
A theoretical approach combining the three memory tiers (DRAM, NVM and SSD) for database systems is discussed by Alexander van Renen \emph{et al.}~\cite{van2018managing}. 
All those research topics focus on a low-level integration, with special highlights on the caching policies of those systems and their performance from within the storage system. Our main focus is in the higher level integration, specifically on how an \emph{active} object store would be able to take advantage of the new memory tier and exploit its byte-addressable nature.

The potential of NVM devices has also been explored within the database management field. For instance, Facebook researchers designed a system using NVM~\cite{eisenman2018reducing} which resulted in a decrease of the DRAM footprint in a production data center environment while maintaining comparable performance; the use case in that article is strictly focused on a MySQL storage engine implementation: MyRocks~\cite{myrocks}. 
Similar results can be obtained with an Online Transaction Processing database system as shown in the work of Liu \emph{et al.}~\cite{liu2021zen}. Those articles demonstrate how a database management system can leverage NVM in order to achieve a performance comparable to DRAM at a fraction of its cost.

\subsection{Dedicated hardware for active features}

The concept of processing data from within its location has also been explored from the hardware standpoint; that is the case for Intelligent Disks (IDISKs~\cite{keeton1998case}) and also discussed by Kannan \emph{et al.}~\cite{kannan2011iostaging, kannan2011cloudio}. That last line of research evaluates the active paradigm --on an architecture that includes additional hardware compute units-- with NVRAM technologies. 
Nider \emph{et al.} discuss \emph{Processing in Storage Class Memory}~\cite{nider2020processing}, an article that shows promising results on this main idea of processing in NVM. The data locality obtained during execution in those articles resembles the one that we will be presenting, but our infrastructure will be based in a \emph{software} storage system, more precisely an \emph{active object store} --which takes advantage of the NVM byte-addressable nature through the \texttt{dax} features.

\section{Object store}

\label{sec:objectstore}

\subsection{Foundations}

The starting point that we are considering is an object store, a software stack capable of storing \emph{objects} (e.g., \cite{openstack-swift, amazon-s3, ceph-web}). The canonical way of using this kind of storage system is through the \emph{get}/\emph{put} interface of the object store. The \emph{get} operation retrieves a byte array --i.e., a previously stored  \emph{object}). The \emph{put} operation sends the object data --again, in the form of byte array-- for the storage system to store it.

From the point of view of the application, the dataset is split onto objects and stored in the storage system. The idea of distributing (splitting) a dataset onto objects is crucial for distributed environments --both from the point of view of distributed storage \emph{and} distributed execution. The distribution mechanisms fall outside the scope of this article, but they are well-known and ubiquitous; we will be following this ``splitting datasets onto objects'' main idea both for the discussion as well as for the evaluation of applications.

In our specific environment, we will be using an object store with additional features, such as the active capabilities and the ability to take advantage of NVM memory. Those aspects are discussed in the following subsections.

\subsection{Active capabilities}

The first goal of our proposed storage system is the availability and exploitation of \emph{active} features. The \emph{active} feature is a mechanism that will enable the execution of code from inside the object store; that code will be able to access, modify and/or process objects residing within the object store.

We will refer to the code being executed in the object store as \emph{active method}. In order to have a flexible and versatile solution, those methods should be application-specific. This implies that there should be some code-shipping mechanism that allows the application developer to send the code for those \emph{active methods} to the storage system space; we will consider a registration mechanism that happens on application initialization. 
The aforementioned strategy is not the only way of achieving our goal, but we believe that doing this \emph{active method} registration on application initialization is a reasonable approach: the \emph{active method} code definition/registration can be application-driven and this process is performed outside the execution critical path --i.e., it does not impact application performance.

During application execution, after the \emph{active methods} are registered, the application can invoke those methods.
This resembles a Remote Procedure Call mechanism; but the gist of this execution (the \emph{active} feature) is that the object store will have the objects directly accessible and thus will be able to access them efficiently and execute the code on them, and then return the result --this, in most cases, avoids the needless serialization, transfer and deserialization of the whole object.

The \emph{active methods} --the code executed by the active storage system-- will typically be a fairly precise and narrow data\hyp{}centric code.
Note that the characteristics of the shipped code are merely a guideline, as the discussion in this article imposes no constraints regarding the complexity and shipping of said code. 
However, it is realistic to assume that code offloaded into the storage systems will be data\hyp{}intensive --the code that will benefit the most from the data locality and active capabilities of the system. 

\subsection{NVM usage}

There are two strong motivations that have driven us to consider supporting NVM devices in our proposed active storage system:

\begin{itemize}
    \item They have much better performance than drives, available at a fraction of the price\hyp{}per\hyp{}byte cost of DRAM.
    This allows the object store to manage big datasets --datasets that would not fit in DRAM-- without incurring in the high performance penalties of storing and accessing them to/from drives.
    \item They are byte\hyp{}addressable, which enables in-place execution 
    (i.e., direct load/store accesses are possible and efficient, without moving data to DRAM). Being able to execute code in-place can boost the data locality; this perfectly combines with the idea of an \emph{active} storage system that is itself improving data locality.
\end{itemize}

The NVM devices that we will be using for the evaluation are Intel Optane DC devices, which have two main operating modes~\cite{dcpmoperatingmodes}: \emph{Memory Mode} and \emph{App Direct}.

\begin{figure}
    \centering
    \subfigure[Memory Mode]{
    	\includegraphics[width=0.46\columnwidth]{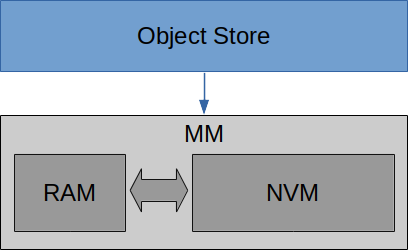}
    	\label{fig:memorymode}
    }
    ~
    \subfigure[App Direct mode]{
        \includegraphics[width=0.46\columnwidth]{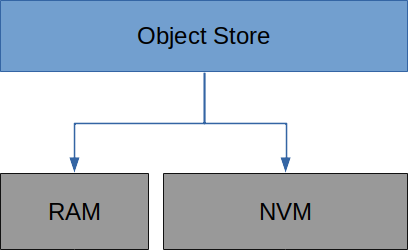}
    	\label{fig:appdirect}
    }
    \caption{Configuration modes for the OptaneDC NVM}
\end{figure}

The first operation mode (called \emph{Memory Mode}, Figure~\ref{fig:memorymode}) presents a big (but volatile) system memory address space, transparently accessible by all software. This is achieved by the aggregation of the Optane DC memory into a single flat memory space. This memory space is transparent to the object store and any software being run in the system. The DRAM of the system is used as cache for this memory space. The hardware will be the one responsible of moving data back and forth between DRAM and NVM. This mode of operation does not require any software adaptation to the OptaneDC devices. 

Note that under Memory Mode configuration the memory space remains volatile in nature; if persistence is desired, it should be managed outside the Memory Mode bound devices.

When the system is configured in \emph{App Direct} (Figure~\ref{fig:appdirect}), the system memory is unchanged and the Optane DC devices memory is persistent and independent from the main system memory. This is the canonical way of using any NVM device: an additional memory tier that can be accessed from the application/middleware. Under this configuration the object store is able to access data both in the DRAM and in the NVM, in a byte-addressable fashion. In order to use that memory space, applications can take advantage of the \emph{Direct Access} (\texttt{dax}) feature which allows applications direct load/store access to persistent memory by memory-mapping files on a persistent memory aware file system~\cite{website_pmdk, nvmprogrammingmodel}.

The assignment of persistent memory into one or the other operation mode is a static parameter and changing the configuration of the Optane DC devices will require a reboot of the machine. It is also possible to segregate the memory into two regions and use a different mode in each one, and thus each region will behave according to its configuration. In this article we will evaluate and discuss each mode separately, which will allow us to characterize them properly.

\subsection{dataClay}

The object store that we will be using for the discussion and evaluation is dataClay~\cite{marti2017dataclay}, a distributed \emph{active} object store. Objects in dataClay can have \emph{active methods}, and the framework enables the remote invocation of such methods --i.e., the framework has the required \emph{active} capabilities that we will be discussing and evaluating.

On top of the \emph{active} built-in capabilities, the programming model of dataClay allows us to integrate the NVM onto the storage system and support the different operation modes aforementioned. This is done through a widely used NVM library: \texttt{pynvm} (more information on the software libraries and versions used is provided in Section~\ref{sec:methodology}).
It is worth noting that several Intel developers contribute to this library, which is an official pmem.io~\cite{github_pynvm} library.

The basic workflow of application development with dataClay starts with the class definition. The semantics closely follows the object-oriented programming paradigm: a class is defined on a supported object oriented language (currently Python or Java) and this class includes attributes and methods. The attributes of the class become the schema of the stored objects, and the methods of the class become the \emph{active methods} that we have mentioned earlier. The attributes of an object can be references to other objects (just as one would expect in an object oriented language) and method execution can access other objects as well, and/or other objects' methods.

\added{
The following snippet shows a basic example on Python of a class definition that could be registered into dataClay:
}

\begin{lstlisting}[name=Class definition example,language={Python},frame=single]
class Block(DataClayObject):
    """
    @ClassField data numpy.ndarray
    @ClassField label str
    """
    @dclayMethod(return_='float')
    def mean(self):
        return self.data.mean()
\end{lstlisting}

\added{
This example shows a very simple \texttt{Block} class with two attributes and an active method.
}

\added{
It is encouraged to perform the registration outside the execution critical path, before the application execution (during the initialization step). In dataClay, this registration is done by a client utility; this utility takes the class definition source code, checks its validity and sends it to an API endpoint that processes the code.
After this registration process, any application will become able to persist and use objects. The invocation of a method on a persisted object results in an \emph{active method} execution --which constitutes the \emph{active} capabilities of the object store.
}

\added{
The following code shows how a Python application may use the previous \texttt{Block} class example during execution.
}

\begin{lstlisting}[name=Class definition example,language={Python},frame=single]
from registered.model import Block
...
b = Block("MyLabel", [1, 2, 3])
b.makePersistent()
m = b.mean()  # Remote active method
\end{lstlisting}

\added{
The transparent behaviour of dataClay is achieved through the use of \emph{stubs}. Similarly to the registration process, the retrieval of stubs is done through the dataClay client utility. The utility will connect to the dataClay API and ask for a class; the endpoint will reply by sending back the stub code of that class.
Those stubs behave identically to the original classes until the moment that \texttt{make\-Per\-sistent} is called, i.e., when the object is persisted into dataClay. From that point on, all attribute accesses and method invocations result in a transparent RPC to dataClay, performed by the dataClay client library.
}

Internally, the dataClay execution subsystem uses the canonical representation of the object according to the object-oriented programming language. This means that a Python-defined class will result in Python objects, allocated within a Python interpreter. When an \emph{active method} is invoked on such objects, the code is executed in the interpreter memory space; there is no need for additional memory copies or data transformations, a fundamental property in order to take advantage of the data locality potential of the system. This property holds true as long as the object is \emph{cached} --i.e., while the object is loaded in the Python interpreter memory space.
Object management within dataClay supports the serialization of objects and their persistence to block storage; this happens automatically on various scenarios, for example when there is memory pressure, or when the storage system shuts down. 
In this article, we are considering an scenario where the node has plenty of NVM memory available --meaning we expect no memory pressure, as the objects can be held in\hyp{}memory/in\hyp{}NVM without issue. This results in executions where all objects maintain that canonical object representation.

In addition to the services which constitute the storage system, dataClay is a library that provides the programming language integration for all those active features to work seamlessly. From the application developer point of view, this means that a call to a method on an object will result in a remote \emph{active method} execution, transparently managed by the client library.

\section{Applications and data structures}

\label{sec:apps}

Our objective is to show how data locality plays an important role on the performance of applications. We have chosen a set of applications that  showcase a diversity of workloads that are representative of several application domains, such as numerical methods, machine learning algorithms, data analytics or big data. The applications we propose are: histogram, \emph{k}-means, matrix addition and matrix multiplication. They are well-known kernels of more complex real\hyp{}life applications.
Not only they are widely used but their behavior is representative of a much larger set of applications; this will be discussed on a per-application basis in the next subsections. 
In this article, and in the context of the chosen applications, we will focus on the impact and potential brought by the active object store, and the general performance attainable with the active features.

For the non-active evaluation, the \emph{method} will be executed in the application itself and the object store will transfer the dataset on demand --just as we would naturally do with an object store, with no active features nor remote code execution of any kind. The division of the dataset and the computation hardware remain the same in both the active and non-active evaluations; see Section~\ref{sec:methodology} for more details on the software and hardware configuration.

\subsection{Overview and comparison}
\label{apps:overview}

\begin{table*}
\begin{center}
\begin{tabular}{rl||c|c|c|S|S|p{1.8cm} }
 \multicolumn{2}{r||}{\textbf{Application}} &
 \multicolumn{1}{|p{1.8cm}|}{\textbf{Data access pattern}} & 
 \multicolumn{1}{|p{1.1cm}|}{\textbf{Object reuse}} &
 \multicolumn{1}{|p{1.1cm}|}{\textbf{Reuse factor}} &
 \multicolumn{1}{|p{1.9cm}|}{\textbf{Computation to data ratio (ms/MB)}} & 
 \multicolumn{1}{|p{1.9cm}|}{\textbf{\emph{Method} comp. index (ms/MB)}} & 
 \multicolumn{1}{|p{1.6cm}}{\textbf{Output size ratio (byte/byte)}}\\
 \hline
 \hline
 \multirow{2}{*}{Histogram} & \emph{(a)} & \multirow{2}{*}{Sequential} & \multirow{2}{*}{No} & \multirow{2}{*}{---} & 7.65 & 7.63 & \multirow{2}{=}{$\sim 10^{-8}$\textsuperscript{\P} \newline 2kB const }\\
 \cline{2-2}\cline{6-7}
 & \emph{(b)} & & & &  7.76 & 7.63 \\
 \hline
 \multirow{2}{*}{\emph{k}-means} & \emph{(a)} & \multirow{2}{*}{Sequential} & \multirow{2}{*}{Yes} & \multirow{2}{*}{10x\textsuperscript{\dag}} & 6.57 & 5.08 & \multirow{2}{=}{$\sim 10^{-6}$\textsuperscript{\P}\newline 80kB const}\\
 \cline{2-2}\cline{6-7}
 & \emph{(b)} & & & & 8.82 & 4.31 \\
 \hline
 \multirow{2}{*}{Matrix add.} & \emph{(a)} & \multirow{2}{*}{Sequential} & \multirow{2}{*}{No} & \multirow{2}{*}{---} & 0.0507 & 0.0457 & \multirow{2}{=}{0.5}\\
 \cline{2-2}\cline{6-7}
 & \emph{(b)} & & & & 0.139 & 0.0107\\
 \hline
 \multirow{2}{*}{Matrix mul.} & \emph{(a)} & \multirow{2}{*}{Non-sequential} & \multirow{2}{*}{Yes} & 6x\textsuperscript{\ddag} & 5.54 & 4.50 & \multirow{2}{=}{0.5}\\
 \cline{2-2}\cline{5-7}
 & \emph{(b)} & & & 42x\textsuperscript{\ddag} & 20.8 & 5.95 \\
\end{tabular}
\end{center}

\footnotesize{\emph{(a)} big objects} \qquad \footnotesize{\emph{(b)} small objects}

\footnotesize{\P{} Given that these algorithms have a constant output size, this value represents the order of magnitude of the ratio for the datasets chosen in this article (both the small and the big datasets). The output size ratios will naturally decrease as the input size increases.}

\footnotesize{\dag{} The data reuse of the \emph{k}-means algorithm will generally depend on the convergence criteria and the dataset. In this article we have fixed the number of iterations to 10, and that is what the evaluation will show.}

\footnotesize{\ddag{} This value reflects the reuse of objects, which is the relevant metric from the object store point of view. The reuse of block submatrices will depend on the shape (in terms of submatrices) of the input matrix. For the matrix multiplication implementation in particular, the reuse ratio will be equal to the number of submatrices in the side of a matrix. I.e., a $42\times$ reuse ratio happens when the input matrices consists of $42\times 42$ submatrices.}

\caption{Comparison on the different complexity and data access patterns for the applications}
\label{tbl:app-comparison}
\end{table*}

The four chosen applications will be used as representatives of a wide scope of applications. The evaluation will consider two different dataset sizes: \emph{small dataset} (which fits in DRAM) and \emph{big dataset} (which does not fit in DRAM). The dataset is divided onto objects, for which we will consider two different sizes: \emph{(a) big objects} and \emph{(b) small objects}. The dataset size has no impact on the object size, only on the number of objects. In the latter --\emph{small objects} scenario--, the objects are shaped to be roughly the size of the CPU cache (about a dozen megabytes) while the former --\emph{big objects}-- are an order of magnitude bigger. 
The object size affects the \emph{method} execution and also has an impact on the granularity of the storage system: smaller objects will result in more \emph{method} invocations, which means more network calls, but with smaller sizes (i.e., less work per object).

The internal structure of the object will naturally vary between applications; for the applications that we will be discussing it will be one of the following:

\begin{itemize}
    \item An array of values --\textbf{Histogram}
    \item An array of multi\hyp{}dimensional points --\textbf{\emph{k}-means}
    \item A submatrix --\textbf{Matrix addition} and \textbf{Matrix multiplication}
\end{itemize}

Before discussing each application individually (on \ref{apps:histogram}, \ref{apps:k-means}, \ref{apps:matrix_addition}, and \ref{apps:matrix_multiplication}), we will first explain the metrics and indicators that will help classify them to better understand their behaviour:

\paragraph{Data access pattern} 

We will differentiate between \emph{sequential} and \emph{non-sequential} access patterns. 
An application is considered to have a \emph{sequential} access pattern if the whole input dataset is accessed in a predetermined ordered sequence starting at the beginning and finishing at the end, without ``jumps'' nor ``holes''.

\paragraph{Object reuse}

From the object store point of view, we will consider that an application has object reuse if during its execution a certain object is accessed more than once. An iterative algorithm will obviously have object reuse, but other kind of applications can also have object reuse.

\paragraph{Reuse factor} 

This factor will be applicable to applications that have \emph{Object reuse}. This factor is equivalent to how many times each object is accessed during an application execution.

\paragraph{Computation to data ratio}

This ratio is obtained by dividing the computation time of a baseline execution --as shown later in section~\ref{sec:evaluation}-- by the input dataset size. The units used are milliseconds by megabytes. This ratio gives insight on the general behaviour of the application. As a rule of thumb, a low ratio will be an indicator of a memory bound application with low execution times. On the same fashion, a high ratio would be a main trait of compute bound applications that have high execution times.

\paragraph{Method computation index}

We know beforehand how many times a certain \emph{method} will be executed along a full application run. The \emph{method} computation index is obtained by dividing the total \emph{method} execution time (extrapolated from the \emph{method} execution time baseline) by the aggregated input size of all the \emph{method} calls. The units used are the same as the ones used in the \emph{computation to data ratio}: milliseconds by megabytes. This index gives insight on the \emph{method} behaviour while also taking into account (implicitly) the object reuse --the input size of all the \emph{method} calls will count with multiplicity the input dataset.

\paragraph{Output size ratio} 

The quantity of write operations is related to the size of the output dataset. The output size ratio is obtained by dividing the input dataset by the output, which gives a first intuition on the relevance of write operations for the application. It is a unitless ratio.

Table~\ref{tbl:app-comparison} contains the results for all the previously explained metrics. The rest of this section contains further discussion on each individual application.

\subsection{Histogram}
\label{apps:histogram}

The first application is a histogram done onto an array of floating point numbers generated randomly following a $F$\hyp{}distribution. The $F$\hyp{}distribution has been used because its shape is easily identifiable, it is asymmetric and non-negative. The intervals of the histogram are 140 predefined bins, spanning from $0$ to infinity, and are non-homogeneous in size. The histogram \emph{method} aggregates the number of points that fall inside a bin, and it returns those values --an array of integers, with as many entries as bins. For a histogram execution the whole input dataset is read once; this means that this application has a \emph{sequential data access pattern} with \emph{no object reuse}.

The general behaviour of the histogram application will resemble a whole family of scientific and big-data applications such as: filtering, max/min finding, linear searches, post\hyp{}processing\ldots{} In general, the \emph{reduce} step (from a \emph{map-reduce} programming model application) will also follow this same memory access pattern.  The basis of our discussion will be an implementation of the histogram application but the results and discussion can be extrapolated to all those other applications --applications which read a big dataset once with no sizeable write operations.

\begin{figure}
\includegraphics[width=\columnwidth]{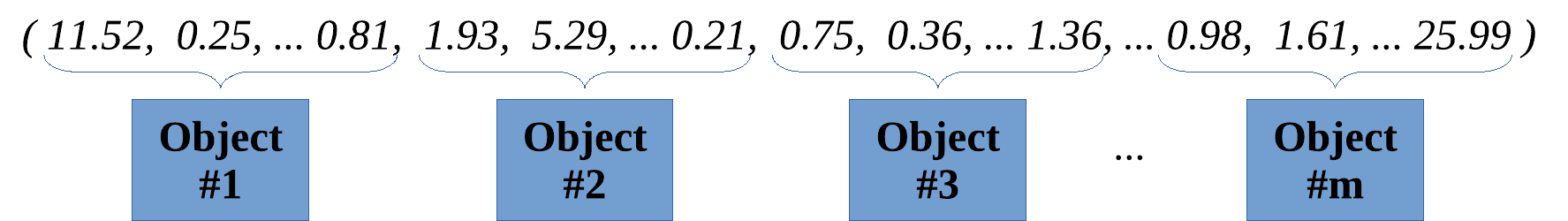}
\caption{Object data structure for histogram input dataset}
\label{fig:datastruct-histogram}
\end{figure}

\paragraph{Data structure} The input dataset is an array of floating point values. This data structure is split into blocks, each block becoming an object in the object store --as shown in Figure~\ref{fig:datastruct-histogram}. Those objects are \verb+numpy+ arrays of floating point values. The output data is a \verb+numpy+ array containing integers --the aggregates associated to the bins. This result is of small size and is returned by the \emph{method} --it is not stored in the object store. The output size is several order of magnitudes smaller than the input, and it is constant in size.

\paragraph{Method characteristics} The code computes the histogram for a single object (a block). This \emph{method} returns the evaluation a partial histogram and --afterwards, outside the \emph{method}-- all the results are merged onto the final histogram.

\subsection{\emph{k}-means clustering}
\label{apps:k-means}

The next application that will be discussed is an implementation of the iterative \emph{standard algorithm} (also called \emph{Lloyd's algorithm}) for the $k$-means clustering method. Given its ubiquity and for the sake of brevity, we will refer to either the method, the algorithm, and the implementation, as simply $k$-means.

The $k$-means is an iterative implementation in which, for each iteration, all the points are read and new center clusters are evaluated. As one can imagine, the memory access pattern for this is completely deterministic and predictable, given that all the points are accessed in a single sweep fashion and only a small quantity of data needs to be written: the cluster centers --new ones are generated for each iteration. This application presents a \emph{sequential data access pattern}, with \emph{object reuse}. The reuse factor will depend on the number of iterations.

The input dataset is a cloud of points generated randomly before starting the application. The number of centers is fixed at 20. Instead of waiting for convergence criteria of said centroids we have fixed the number of iterations to 10 --i.e., the \emph{reuse factor} is equal to 10. This configuration setup does not change the computational complexity, nor does it have any impact on the execution time of a single iteration. Obviously, a realistic dataset may need more iterations before reaching convergence and that number of iterations would depend on the initial centers. However, 10 iterations is enough to showcase the data reuse phenomenon of the algorithm and ensures that the results are repeatable and can be extrapolated to real datasets.

We have chosen the \emph{k}-means clustering algorithm as a representative of machine learning algorithms that perform several sequential reads of the input data. Of course this general behaviour is not unique to the \emph{k}-means and we can find other scientific and machine learning applications with similar patterns (genetic algorithms, temporal simulations\ldots). The results and discussion in this article can be extrapolated to any of those applications that share this same memory access pattern --a big dataset that is read, whole, several times during the application execution.

\begin{figure}
\centering
\includegraphics[width=0.7\columnwidth]{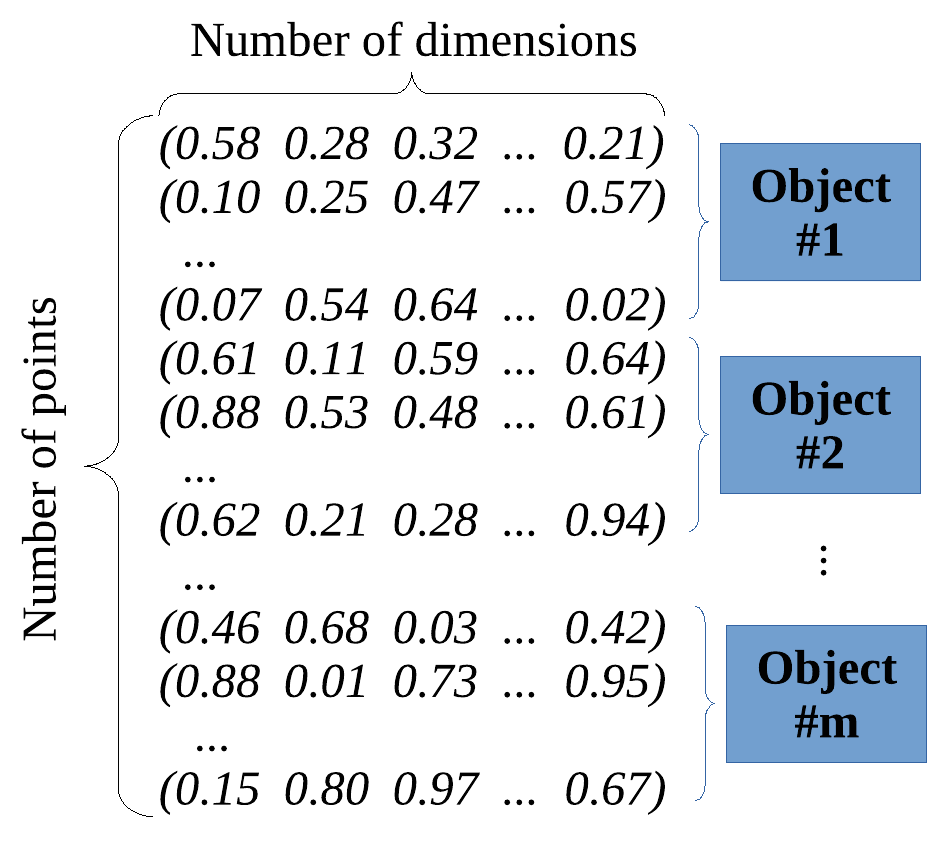}
\caption{Object data structure for \emph{k}-means input dataset}
\label{fig:datastruct-kmeans}
\end{figure}

\paragraph{Data structure} The input dataset representation is an array of $n$-dimensional points. This data structure is split into blocks, each block becoming an object in the object store --as shown in Figure~\ref{fig:datastruct-kmeans}. Those objects are \verb+numpy+ matrices, each of which represents a set of $n$-dimensional points. The output data is itself a \verb+numpy+ matrix, representing the centroids (which are $n$-dimensional points). This centroids structure is quite small as the number of centroids will be orders of magnitude smaller than the number of points, and thus the \emph{method} will return the centroids data structure by value --they will not be stored in the object store.

\paragraph{Method characteristics} The \emph{method} part of the algorithm receives the last iteration centroids as an input. This function is able to perform categorization for the points in a single object (the current block) and also evaluate a partial summation for its points given the centroids. All the partial summations can then be processed --outside the \emph{method}-- in order to obtain the centroids for the next iteration. Note that both centroids and partial summations are small data structures; their size is related to the number of centroids which is orders of magnitude smaller than the number of points.

\subsection{Matrix addition}
\label{apps:matrix_addition}

The matrix addition implementation will use randomly generated matrices (two $n \times n$ square matrices) and perform their addition (resulting in a third $n \times n$ square matrix). The complexity of the matrix addition operation is $O(n^2)$.

The data access pattern for the addition operation is predictable and deterministic: a single sweep is done to both input matrices; within this sweep the data is written to an output matrix, without reusing the input blocks. The application presents a \emph{sequential data access pattern}, with \emph{no object reuse}.

The memory access pattern of this application can be seen in a lot of different applications, meaning that the discussion and results obtained for this application can be extrapolated to those other ones. For example, \emph{map} functions (from \emph{map-reduce} programming model applications) fall within this spectrum, as well as most data transformations kernels. All those applications are implemented by being applied to a big dataset, sequentially, and yielding an output dataset, which is exactly the general behaviour of the matrix addition.

\begin{figure}
\includegraphics[width=\columnwidth]{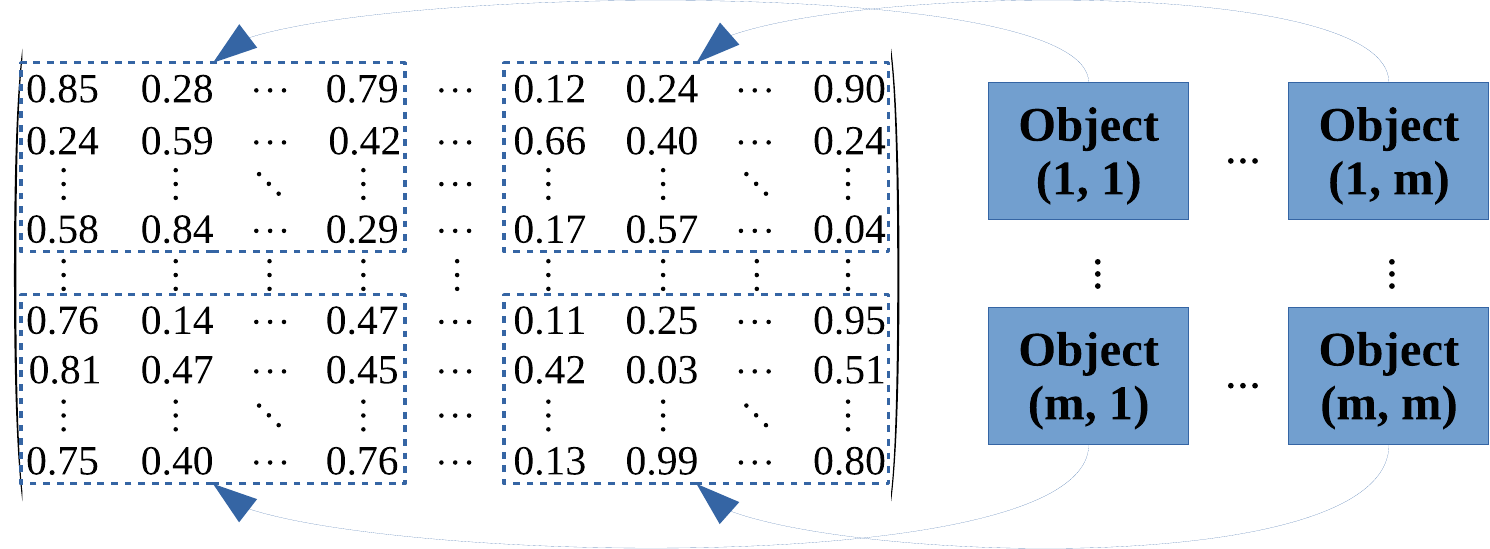}
\caption{Object data structure for matrix data structures}
\label{fig:datastruct-matrix}
\end{figure}

\paragraph{Data structure} The input dataset are two square $n \times n$ matrices. The output dataset is a single $n \times n$ matrix. Each matrix is represented as a two-dimensional array of submatrices --each submatrix is $k \times k$. Those submatrices are stored as objects in the object store --as shown in Figure~\ref{fig:datastruct-matrix}. The size of the output matrix equals the size of either of the input matrices, meaning that the \emph{output size ratio} is equal to one half.

\paragraph{Method characteristics} The addition of two objects (two $k \times k$ submatrices) is what the \emph{method} computes. The output of that function is itself a new object (a submatrix) that may or may not be put into the object store --i.e., persisted and/or put in NVM. We will specifically discuss the placement of this output object in the evaluation.

\subsection{Matrix multiplication}
\label{apps:matrix_multiplication}

The matrix multiplication will use (just as in the matrix addition) randomly generated $n \times n$ matrices. We will consider the iterative multiplication implementation with $O(n^3)$ complexity. All previous applications have a computational complexity that matches their storage requirements. But that is not satisfied by the matrix multiplication: the computational complexity ($O(n^3)$) has a substantially larger growth than its storage ($O(n^2)$). That difference between storage and computation means that the storage access overhead becomes less and less important (performance\hyp{}wise) when the dataset is increased.

The data access pattern for matrix multiplication is more complex than the matrix addition. In the discussed implementation, which uses ``block matrices'', the computation of the result is done by iteration of the output blocks. Each output block requires a certain quantity of multiplications and additions of input blocks. Once those multiplications and additions are finished, the output block can be stored and it won't be reused again. Note that there is reuse of the input blocks, as any block in the input matrices will be used to compute more than one block of the output matrix. This makes this application have much more data reuse than previous ones; and even though the reuse of data follows a deterministic pattern, it is not a trivial one --e.g., it is not multiple full sweeps as we could observe in the $k$-means. This application is a representative of \emph{non-sequential data access pattern} with \emph{object reuse} --as well as having a \emph{method} implementation with \emph{data reuse}.

We have chosen the matrix multiplication as an iconic and well-known numerical kernel, although we can find other matrix algorithms that will follow similar memory access pattern on input and output matrices --for example, matrix decomposition algorithms. One can also expect this pattern of non-sequential reads onto big datasets in certain applications like mesh-based simulations, data analytics, and others.

\paragraph{Data structure} Same as in \ref{apps:matrix_addition}~\emph{Matrix addition}.

\paragraph{Method characteristics} The \emph{method} is responsible of performing a \emph{multiply-accumulate operation}. This is used with the (initially zero-initialized) output objects (submatrices) in order to perform the matrix row-by-column multiplications and additions. Using multiply-accumulate mechanisms --which are sometimes called \emph{fused mutiply-add} or \emph{FMA} in low-level jargon-- for matrix multiplication is a well-known approach and it is appropriate for a \emph{block matrix multiplication} algorithm.

\section{Methodology}
\label{sec:methodology}

\subsection{Hardware platform}

\label{methodology:hwplatform}

All the experiments will be done in a system equipped with Optane DC Persistent Memory Modules. The technical specs of that node are the following:

\begin{itemize}
    \item 2$\times$Intel\textregistered{} Xeon\textregistered{} Platinum 8260L CPU @~2.40GHz
    \item 192GB (12$\times$16GB, 2666MHz) of DRAM 
    \item 6TB (12$\times$512GB modules) of Intel\textregistered{} Optane\texttrademark{} DC Persistent Memory 
\end{itemize}

This machine contains a total of 48 ($2 \times 24$) usable cores. The machine supports both the \emph{Memory Mode} and \emph{App Direct} execution modes; a reboot is required in order for its mode to be changed.

All the experiments will be evaluated with the same single machine, both while using dataClay (active object store) and while using DAOS (non-active object store). The executions will be done sequentially, although the numerical libraries will be taking advantage of the multiple cores through their internal multithreading capabilities. The availability of two CPU sockets allows us to isolate the application and the object store by placing them (by binding processes) to different sockets.

\added{
If we look into the hardware cost and compare the cost per gigabyte of DRAM versus NVM we can expect the latter to be lower; for typical builds, Intel reports savings of up to 30\%\cite{dcpmbrief}. Persistent Memory also allows a system to dramatically increase the total available memory. For starters, one can easily find 512GB Persistent Memory DIMM modules. Moreover, the price of a DRAM DIMM can be 5 to 10 times higher than a same\hyp{}sized Optane Persistent Memory DIMM\cite{tomshardwareoptanedimm}.
}

\added{
Moreover, the evolution of storage technologies has always led to increasing their performance, enlarging their capacity, and becoming more affordable. This happened with consumer hard drives and the pattern repeated with the SSD technology. It is natural to expect a similar evolution with NVM technology.
}

\added{
In this article all datasets will be smaller than the available Persistent Memory. There is no technical limitation behind this decision, as the object store is designed to use the disk when required. However, doing so will incur in staging operations --just as any storage system would. Those disk staging operations are completely orthogonal to the active features and to the evaluation focus.
}

\subsection{Software libraries and deployment}

\added{
The dataClay framework has been deployed following standard installation and configuration procedures, and its Python bindings configured in the system. 
We will follow a common ephemeral strategy regarding deployment and usage of dataClay framework; in our context, the software stack is deployed in the same node as the application and they share computation and storage resources (meaning CPU and memory). An ephemeral configuration means that the storage system is not shared amongst other applications, and that reduces the expected interference problems. The application and the storage systems do share resources, but that results in a fair comparison: application is executed by the same total available resources, whether active features are used or not.
}

%The deployment lifecycle follows the same pattern as the application. Or, in other words, dataClay is deployed as an \emph{ephemeral} active object store. This means that there could be contention between the application and dataClay --although that is kept at a minimum by using the two available sockets, see previous subsection~\ref{methodology:hwplatform}-- but no external interference is expected. This configuration allows the object store to greedily take advantage of all computing resources for its active features.

Both \emph{Memory Mode} and \emph{App Direct} are supported and evaluated. Under the \emph{Memory Mode} configuration, dataClay requires no changes given the transparent nature of the \emph{MM}.

For the \emph{App Direct} configuration, certain code injections are required. The proposed implementation uses the PMDK~\cite{github_pmdk, website_pmdk} libraries (version 1.5), and its \emph{pynvm} Python bindings~\cite{github_pynvm} of PMDK (version: 0.3.1). The numerical structures used in the applications are \verb+numpy+ arrays, represented as contiguous in-memory buffers --their canonical representation. Generally, any data structure could be used as long as its in-memory representation was known --e.g., based on contiguous memory buffers. We are using a recent (2020) version of Intel Distribution for Python~\cite{inteldistributionforpython}.

The non-active storage system used is DAOS~\cite{daos-web}, which is a state of the art object store specifically designed by Intel for non-volatile memories such as NVMe and Optane DC. The DAOS software stack has proven itself to have a very high performance, as shown in the IO500 ranking results~\cite{io500}, where --at the time of writing this document-- DAOS has the highest score.

This DAOS stack has been deployed following the standard installation and set up procedures. In order to reduce external interference to its minimum for the benchmarks, the DAOS software has been configured to be using the Optane DC Persistent Memory and no SSD nor replicas. The DAOS stack will be used in the \emph{App Direct} configuration, which is the mode intended for DAOS.

\section{Evaluation}
\label{sec:evaluation}

In order to discuss the potential of an active object store --compared to a non-active storage system-- and the implications that data locality can have on the application execution performance we will be evaluating the four applications explained before (Histogram, \emph{k}-means, Matrix Addition and Matrix Multiplication) under different configurations.

First and foremost, we will be focusing on the \emph{active} behaviour of dataClay (the object store) while comparing it to a non-active object store (DAOS). The evaluation will take into account the different configuration alternatives of the Optane DC --the available NVM devices.

With all that in mind, we will be evaluating two different dataset sizes. First, a \emph{small dataset}, where the whole data structures (both input and output) are smaller than the available DRAM. The other one, the \emph{big dataset}, is dimensioned to be at least twice as big as the system memory, which ensures that the access to the NVM is realistic --smaller datasets may take advantage of cache through the faster DRAM, which would result in over-optimistic execution times. For each dataset size we will evaluate two different object sizes --as discussed in Section~\ref{apps:overview}. The object size does not impact the total dataset size; big objects will result in fewer \emph{active method} invocations but with a higher cost (due to the higher data per object).

The four different setups (three active ones with dataClay with different memory configurations, and one non-active one with DAOS) are the following:
\begin{description}
\item[dC DRAM] [dataClay, Active] Consists on an application execution where the object store holds all the dataset in DRAM. This can only be evaluated for the \emph{small dataset} as, by design, the \emph{big dataset} does not fit in DRAM. The application will invoke the \emph{active method} within dataClay. The NVM is not used in this scenario. The objective of this configuration is to evaluate the active object store with the fastest memory.
\item[dC NVM] [dataClay, Active] In this configuration, the input dataset resides in the NVM. The application will invoke the \emph{active method} within dataClay. Said \emph{active method} will perform in-place execution. The objective of this configuration is to evaluate the impact of the NVM performance and the potential of in-place execution on NVM.
\item[dC MM] [dataClay, Active] When the \emph{Memory Mode} is configured, the object store holds the dataset in the flat memory space of system memory --that combines the NVM as main memory plus DRAM as a cache layer in a transparent fashion. The application will invoke the \emph{active method} within dataClay. The objective of this configuration is for comparison when the NVM memory space is managed by the hardware with a single memory space --as oposed to the previous configuration where the DRAM and NVM memory spaces are separated and managed at the storage system level.
\item[DAOS] [DAOS, Non-active] The DAOS object store is used as the baseline non-active storage system. In this configuration the dataset is held in NVM by the storage stack. The application will retrieve the objects from the object store and execute the (non-active) \emph{method} within the application space (in DRAM).
\end{description}

\subsection{Histogram}
\label{evaluation:histogram}

\begin{figure}
\includegraphics[width=\columnwidth]{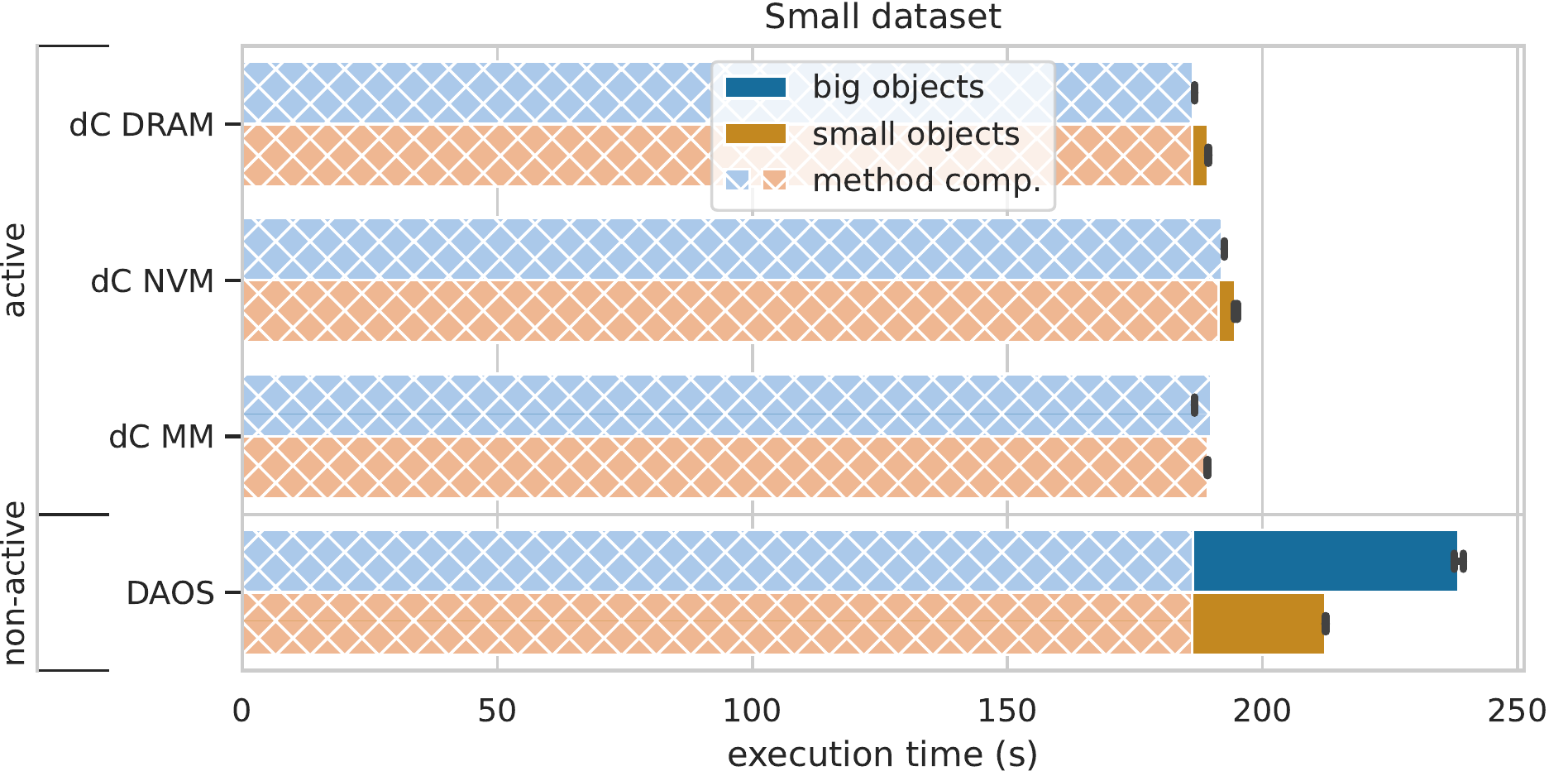}
\caption{Execution times of the histogram application for a small dataset ($n = 3.2 \times 10^9$). \boilerplateappsmall}
\label{fig:histogram-small}
\end{figure}

\begin{figure}
\includegraphics[width=\columnwidth]{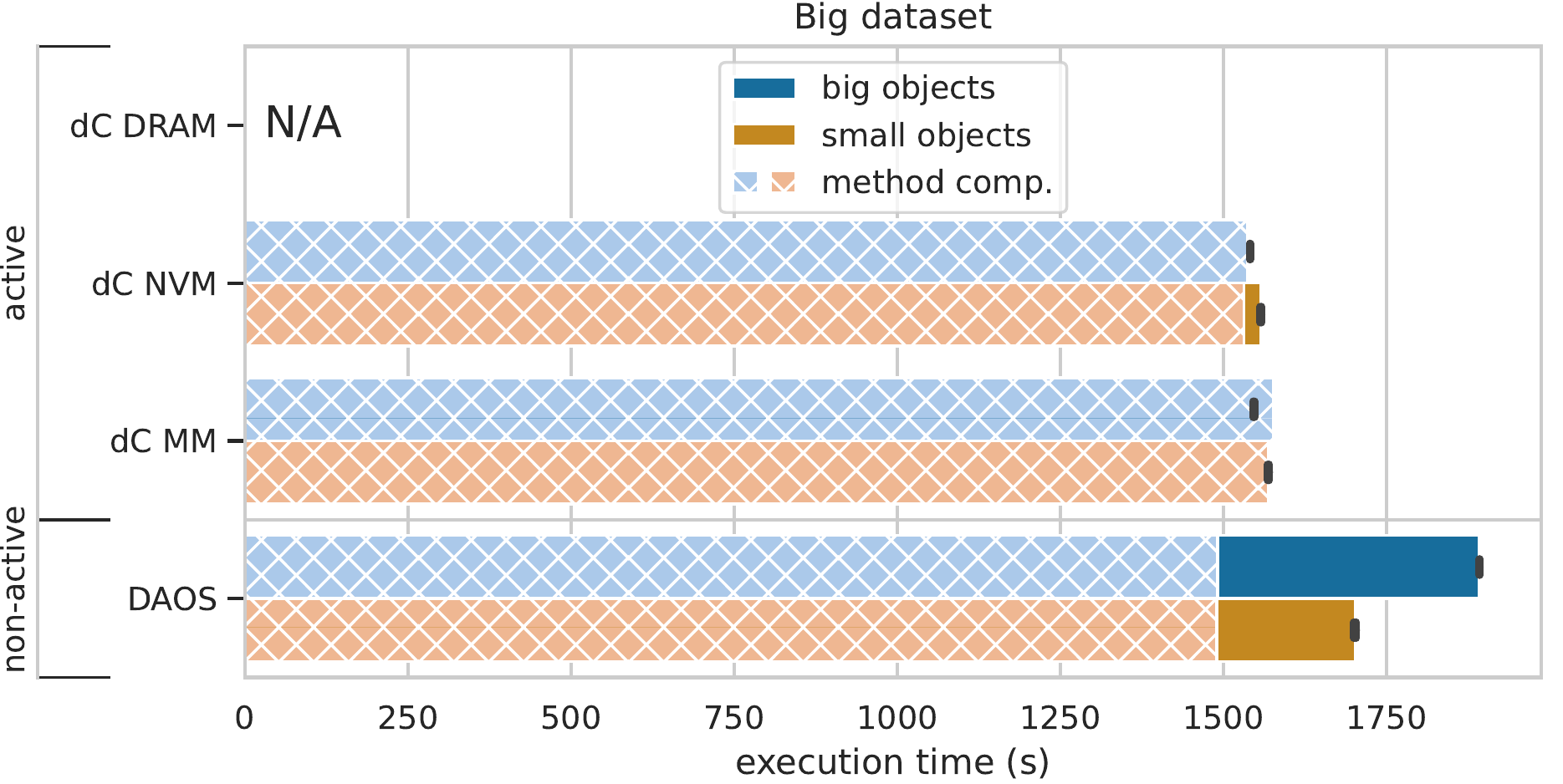}
\caption{Execution times of the histogram application for a big dataset ($n = 25.6 \times 10^9$). \boilerplateappbig}
\label{fig:histogram-big}
\end{figure}

The execution times of the Histogram application can be seen in the following figures: Figure~\ref{fig:histogram-small} shows the execution times for the \emph{small dataset} and Figure~\ref{fig:histogram-big} shows times for the \emph{big dataset}. We will start by looking at the total execution times (the whole bar, ignoring the lighter hatched portion for now) for each configuration --on upcoming paragraphs we will discuss about the meaning and insights given by the \emph{method} computational portion.

All the active execution times, regardless of the dataset size or the object size, are roughly similar --within a 5\% between one another. We can observe that the non-active object store execution (DAOS) is from a 10\% up to a 20\% slower. This shows the overhead caused by the communication --memory transfers and data copies between the non-active object store and the application-- in comparison to the computation. A similar pattern would be shown by any similar application with no data reuse and high computation to data ratio.

It is important to note that, when comparing the non-active DAOS executions, using big objects is slower than using small objects. This shows that the performance toll for the non-active case is not a consequence of the number of RPC calls. In this evaluation, small objects are 8 times smaller than big objects which result in 8 times more RPC. It may seem natural to expect better performance with less RPC calls. Instead, the cost of object transfers --required for non-active execution and the responsibility of the object store-- increases with the object size, which results in slower execution times for bigger objects. This happens under the DAOS non-active executions for both dataset sizes.

\begin{figure}
\centering
\includegraphics[width=\columnwidth]{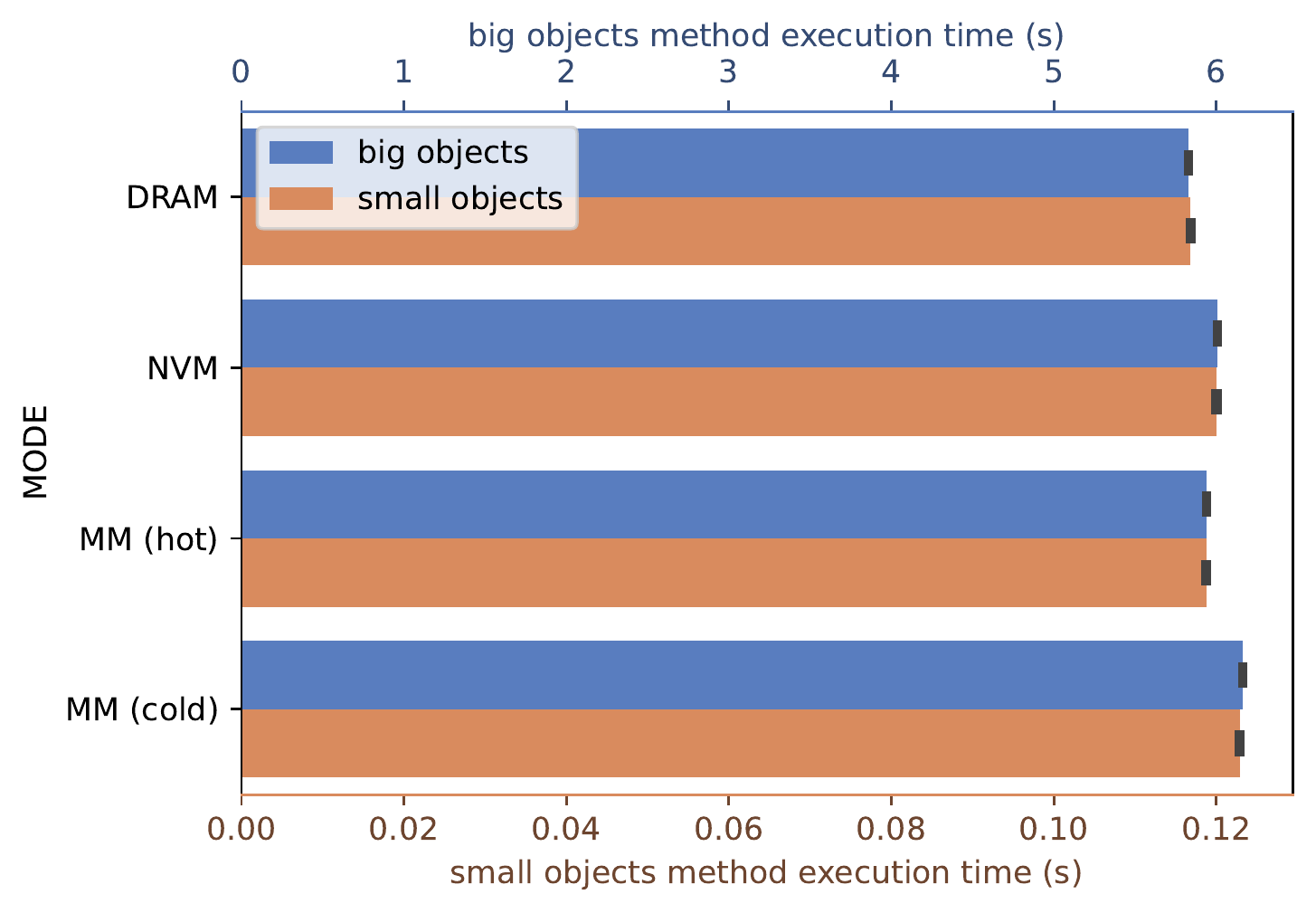}
\caption{Execution times for the histogram \emph{method}. \boilerplatemethod}
\label{fig:histogram-method}
\end{figure}

Upon closer inspection, it is surprising to observe how close all the active configuration execution times are amongst them. In order to shed some light on this behaviour we have included an evaluation of the execution time of the \emph{method} itself --the piece of code that is shipped to the active object store.
Figure~\ref{fig:histogram-method} shows the execution times of the \emph{method} itself when this piece of code is executed either in DRAM, in the NVM, or in the \emph{Memory Mode} OptaneDC configuration. Given that the \emph{Memory Mode} acts as a transparent cache, we have evaluated this last mode in both a hot and a cold configuration. 
Note how for the method execution evaluation under \emph{Memory Mode} we are able to control the environment and consider both \emph{hot} and \emph{cold} execution times.
For the full application execution we are considering the \emph{hot} execution time for the small dataset, as we expect everything to fit in DRAM and thus \emph{hot}. 
The \emph{cold} execution time is used as the estimator for the big dataset; it serves as a worst-case scenario, assuming that all accesses to the data miss the DRAM and go to NVM. The lighter hatched bars in Figure~\ref{fig:histogram-small} and Figure~\ref{fig:histogram-big} show the portion of execution time estimated to be due to the method execution.

We have previously shown (see Table \ref{tbl:app-comparison}) that the histogram \emph{method} has high computation times (high \emph{method} computation index). This means that, for the active execution modes, almost all the execution time comes from the \emph{method} itself and thus the memory configuration has a small impact on the overall performance of the application.

\paragraph{Insights}
We have seen how the active configuration --and the data locality it implies-- introduces a 10\% to 20\% benefit to the overall execution times. This application showcases an scenario with no data reuse and a high \emph{method} computation index. Under those assumptions, the application has low sensitivity to the exact memory configuration, meaning similar performance for all NVM modes tested.

\subsection{\emph{k}-means}
\label{evaluation:kmeans}

\begin{figure}
\includegraphics[width=\columnwidth]{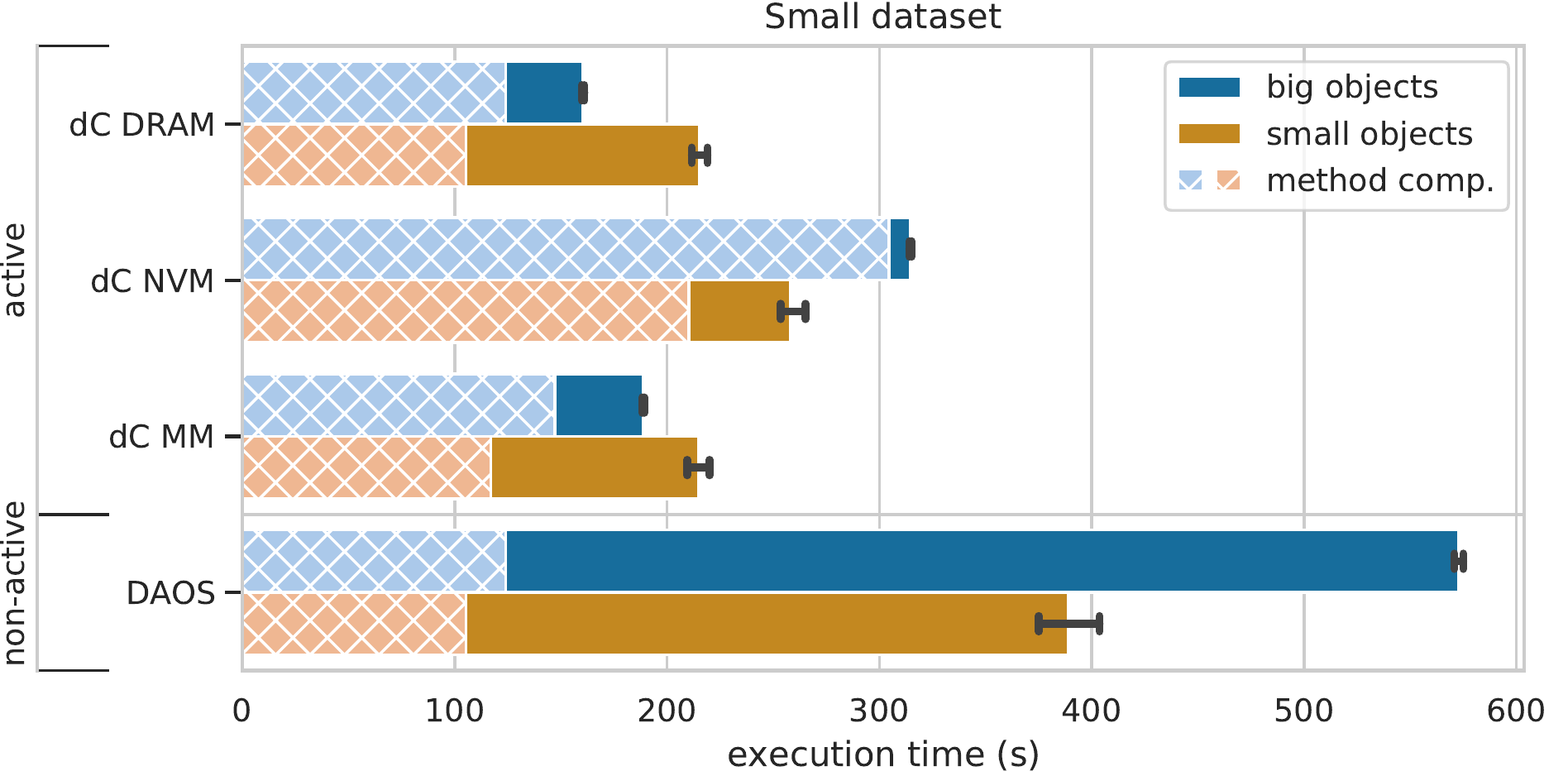}
\caption{Execution times of the \emph{k}-means application for a small dataset ($n = 6.4 \times 10^6$).\boilerplateappsmall{}}
\label{fig:kmeans-small}
\end{figure}

\begin{figure}
\includegraphics[width=\columnwidth]{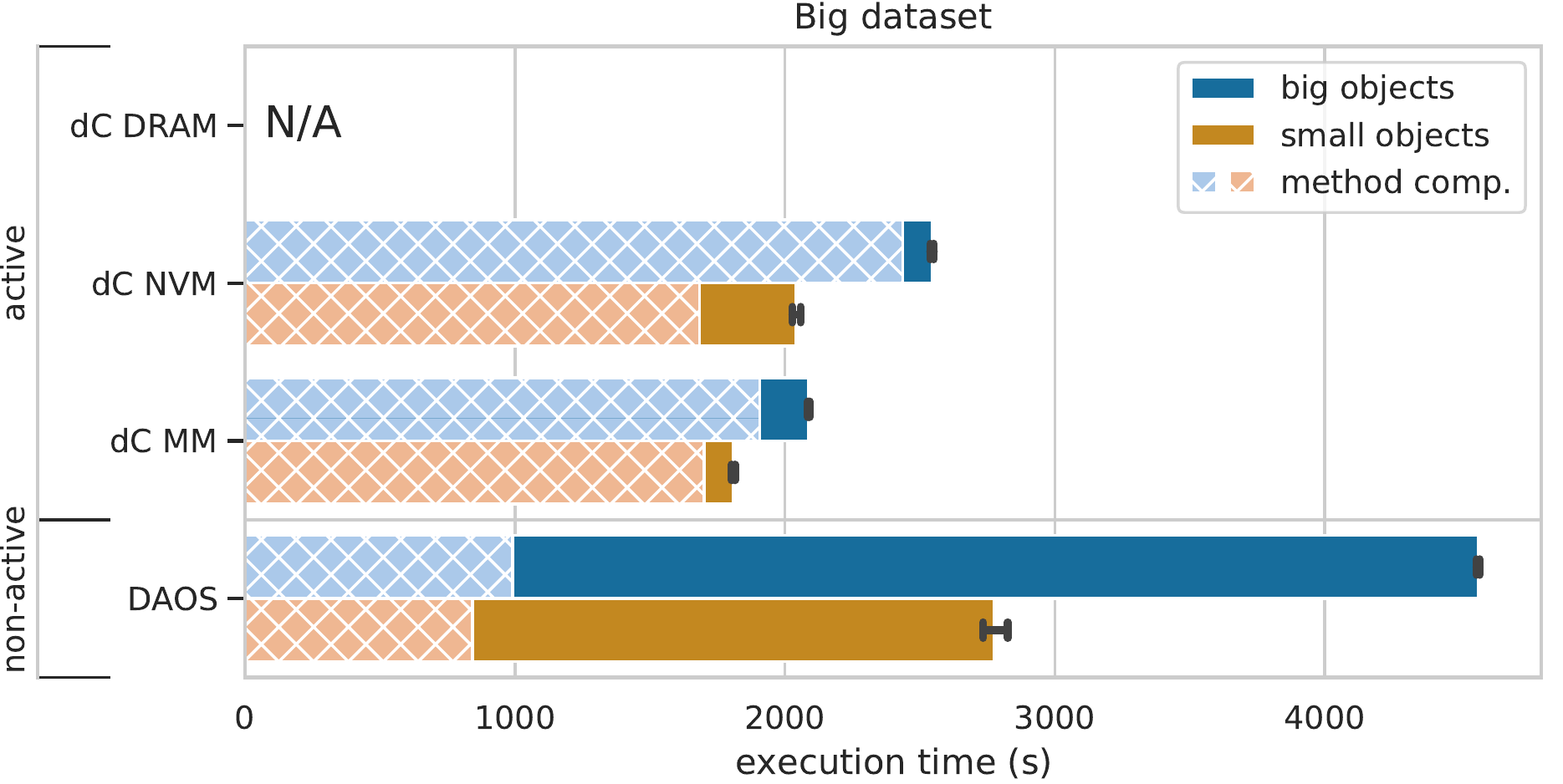}
\caption{Execution times of the \emph{k}-means application for a big dataset ($n = 51.2 \times 10^6$).\boilerplateappbig{}}
\label{fig:kmeans-big}
\end{figure}

The execution times for the \emph{k}-means application can be seen in Figure~\ref{fig:kmeans-small} (for the small dataset) and Figure~\ref{fig:kmeans-big} (for the big dataset). At a first glance we can see how, once again, there is a clear benefit on using an active storage system when compared to the non-active object store. The non-active overhead goes from 30\% --when comparing with the slowest active configuration-- up to a 300\%. This increased gap between the active and non-active setups is greater than on the previous application mainly due to the data reuse of the application --all data is reused for each iteration for a total of 10 iterations. This showcases how, in relative terms under this scenario, the computation time is less relevant and the time consumed by data transfers is more prominent.
However, it is also apparent that this application has higher sensitivity to the object size as well as the memory configuration, as shown by the higher spread on execution times across the different configurations.

The \emph{k}-means \emph{method} being used --mainly, the pairwise distances evaluation as implemented in the \verb+numpy+ library-- has a certain quirk: it reads input data several times. That means that the \emph{method} implementation is performing multiple read operations onto the same object; that is a consequence of \verb+numpy+'s numerical implementation.

\begin{figure}
\centering
\includegraphics[width=\columnwidth]{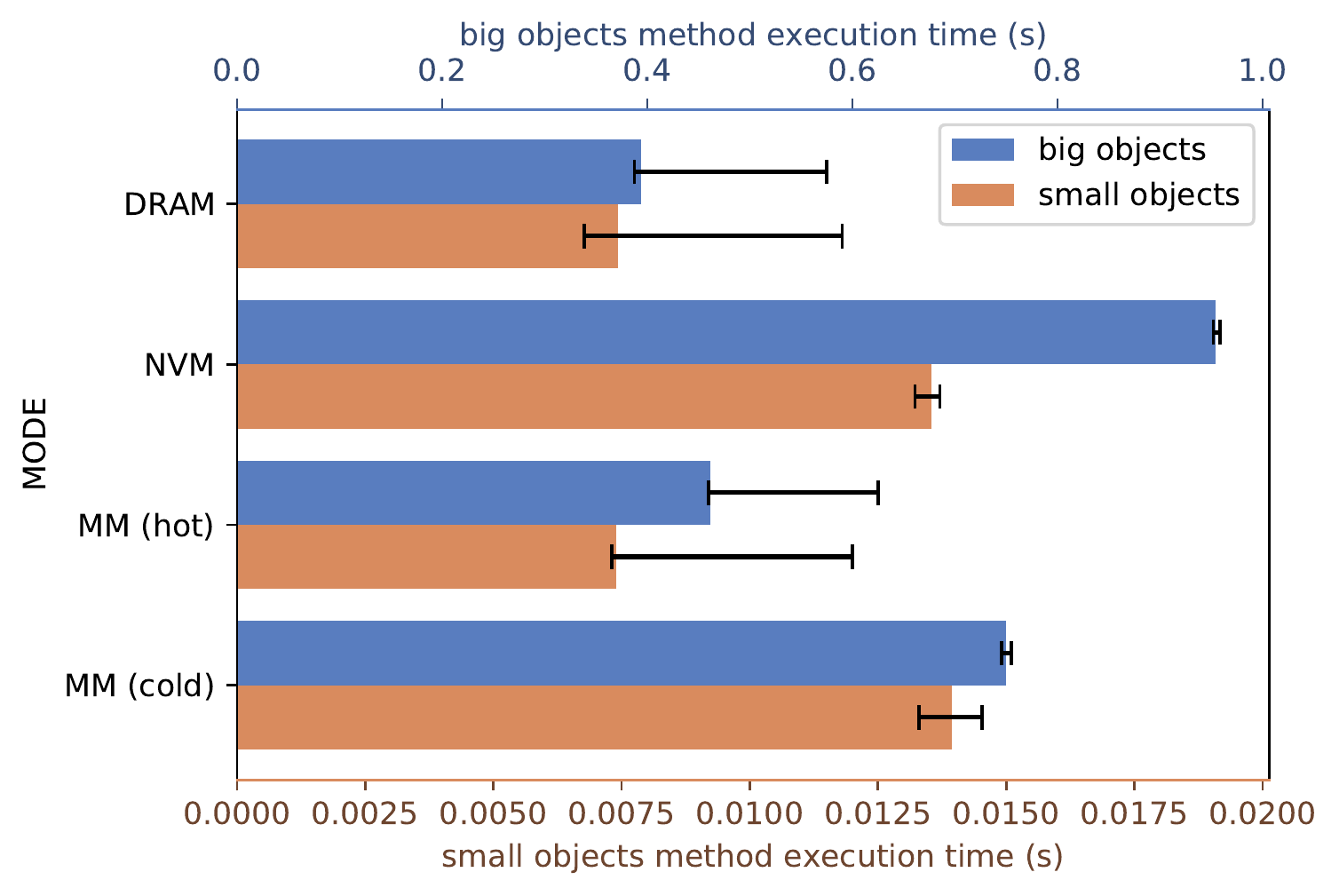}
\caption{Execution times for the \emph{k}-means \emph{method}.\boilerplatemethod}
\label{fig:kmeans-method}
\end{figure}

For a better understanding of the \emph{method} behaviour we have included the execution times in Figure~\ref{fig:kmeans-method}. The high execution times on the NVM memory configuration --i.e., when data is computed in-place in the NVM-- suggest that the data reuse in the \emph{method} is amplifying the difference of speed between the DRAM memory tier and the NVM one.
The implementation sensitivity to memory speed also explains the high variability amongst DRAM-based executions --\emph{DRAM} and \emph{MM (hot)}. 

\paragraph{Insights}
This application shows how algorithms that iterate over the input dataset multiple times --or, more generally, that have certain reuse along their execution-- will obtain great benefits from being executed in an active environment. This is the consequence of the repeated access to the data, which magnifies the gains introduced by the data locality of the \emph{active} object store.
\added{This application also reminds us the relevance of the numerical low-level implementation; the multiple read operations done at the numpy library level has a huge impact on execution times. Further improvements would require low\hyp{}level knowledge on the numerical library implementation and the development of specific strategies for the application --something that only makes sense for heavily used microkernels which warrants the resource investment.}

\subsection{Matrix Addition}
\label{evaluation:matsum}

\begin{figure}
\includegraphics[width=\columnwidth]{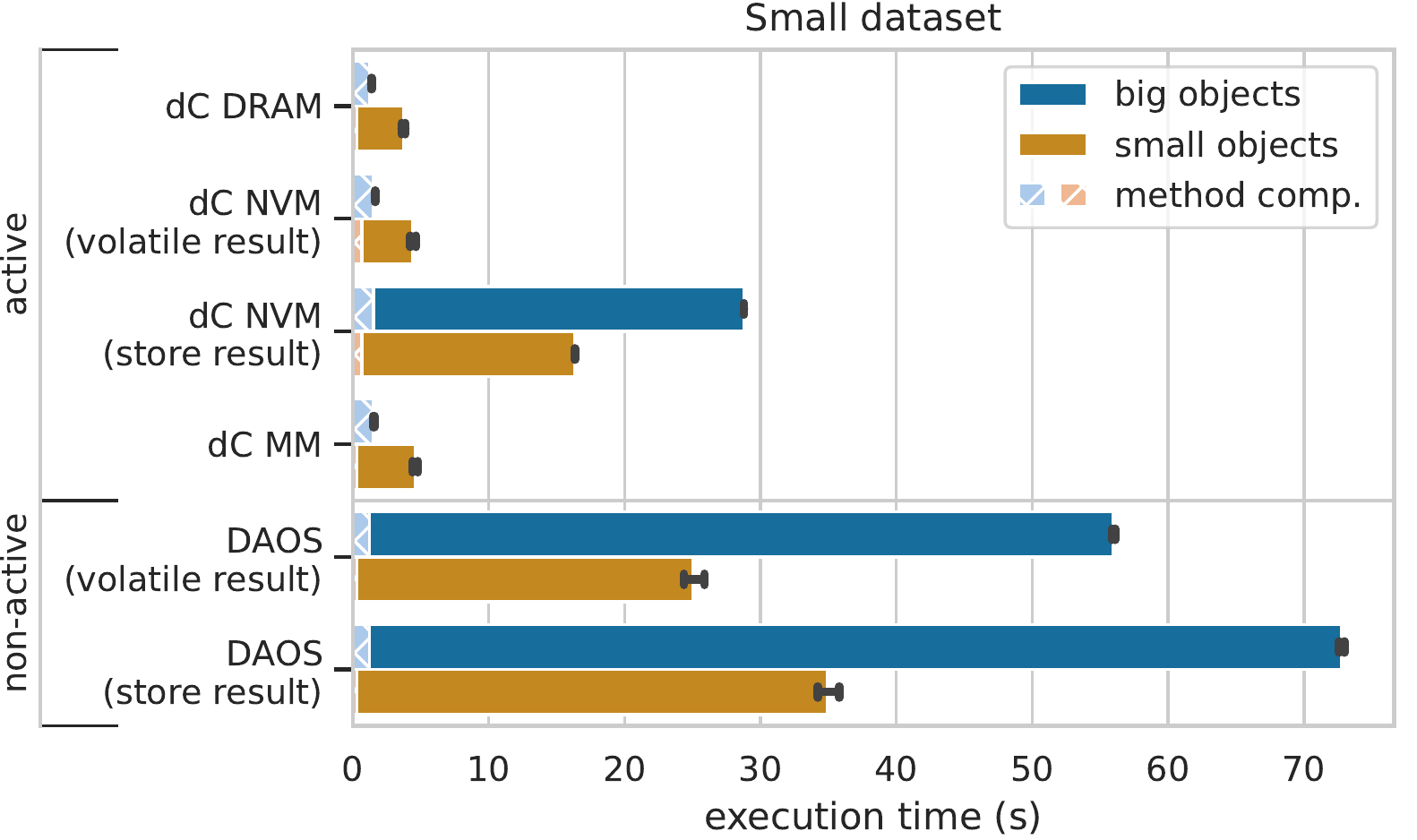}
\caption{Execution times of the matrix addition application for a small dataset ($n = 42000$).\boilerplateappsmall{} The result may be held in DRAM (\emph{volatile result} executions) or stored in the object store (\emph{store result} executions).}
\label{fig:matsum-small}
\end{figure}

\begin{figure}
\includegraphics[width=\columnwidth]{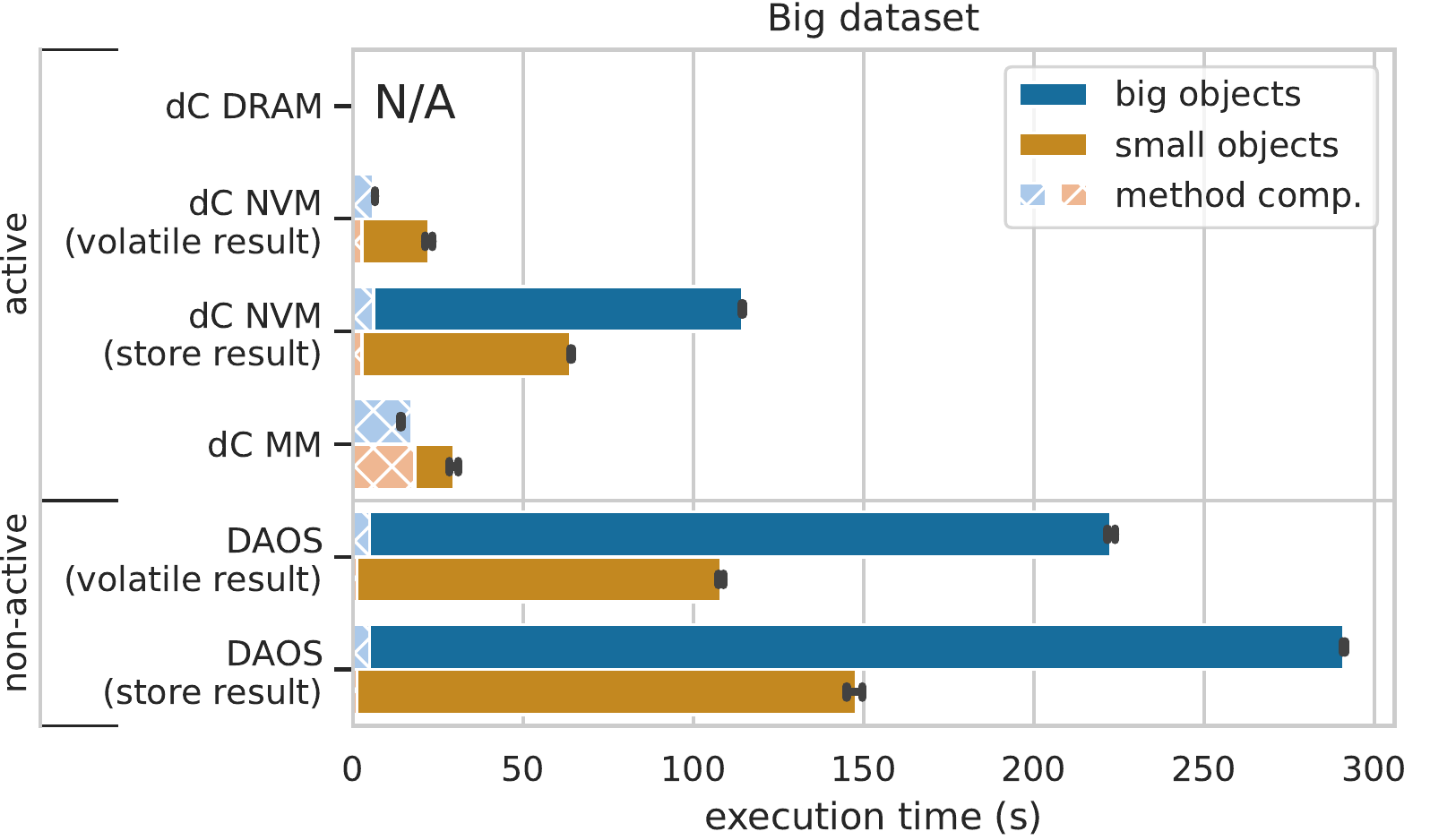}
\caption{Execution times of the matrix addition application for a big dataset ($n = 84000$).\boilerplateappbig{} The result may be held in DRAM (\emph{volatile result} executions) or stored in the object store (\emph{store result} executions).}
\label{fig:matsum-big}
\end{figure}

The next application used for the evaluation is the \emph{Matrix addition}. The execution times for the application when run with a small dataset and a big dataset can be seen respectively in Figure~\ref{fig:matsum-small} and Figure~\ref{fig:matsum-big}.

This is the first application that has a sizeable output, so a distinction is being made on both \emph{DAOS} and \emph{dC NVM}: the result may be stored in the object store (\emph{store result}) or kept as volatile (in the DRAM). This distinction is not applicable to neither the DRAM executions nor the MM ones, as neither of those modes have different addressable memory spaces. Also note that having the output result in DRAM is only possible if the resulting data structure is smaller than the available DRAM.

The decisions on where to store the result will be driven by the amount and availability of system memory --if there is no shortage of memory, then the object store could always choose the fastest memory available, but there will typically be restrictions in-place: memory footprint may be restricted or there may exist certain prioritization.

We will be doing a static decision for the volatile vs store. More complex applications would require additional thoughts related to where to place result data structures. The decision can be entirely developer driven, it can be decided by the software stack, or it can be somewhat in-between by including annotations and reacting to them with runtime information.

The active configurations have a much better performance than the non-active ones, specially when the result is stored in DRAM (\emph{volatile result}) --that active configuration is 15 times faster than the best non-active one.
Given that the matrix addition is a memory-bound application, this result could be expected: avoiding memory transfers is greatly beneficial to the performance, something that the data locality exhibited by the active object store is able to achieve. Moreover, having the result in DRAM while accessing the input dataset through the NVM is a great way to optimize the available bandwidth: the execution is performing all read operations from the NVM while performing write operations to DRAM, avoiding the slower write operations on the NVM and avoiding contention in both memories. That is why this execution mode (\emph{dC NVM (volatile result)}) has the best performance.

\begin{figure}
\centering
\includegraphics[width=\columnwidth]{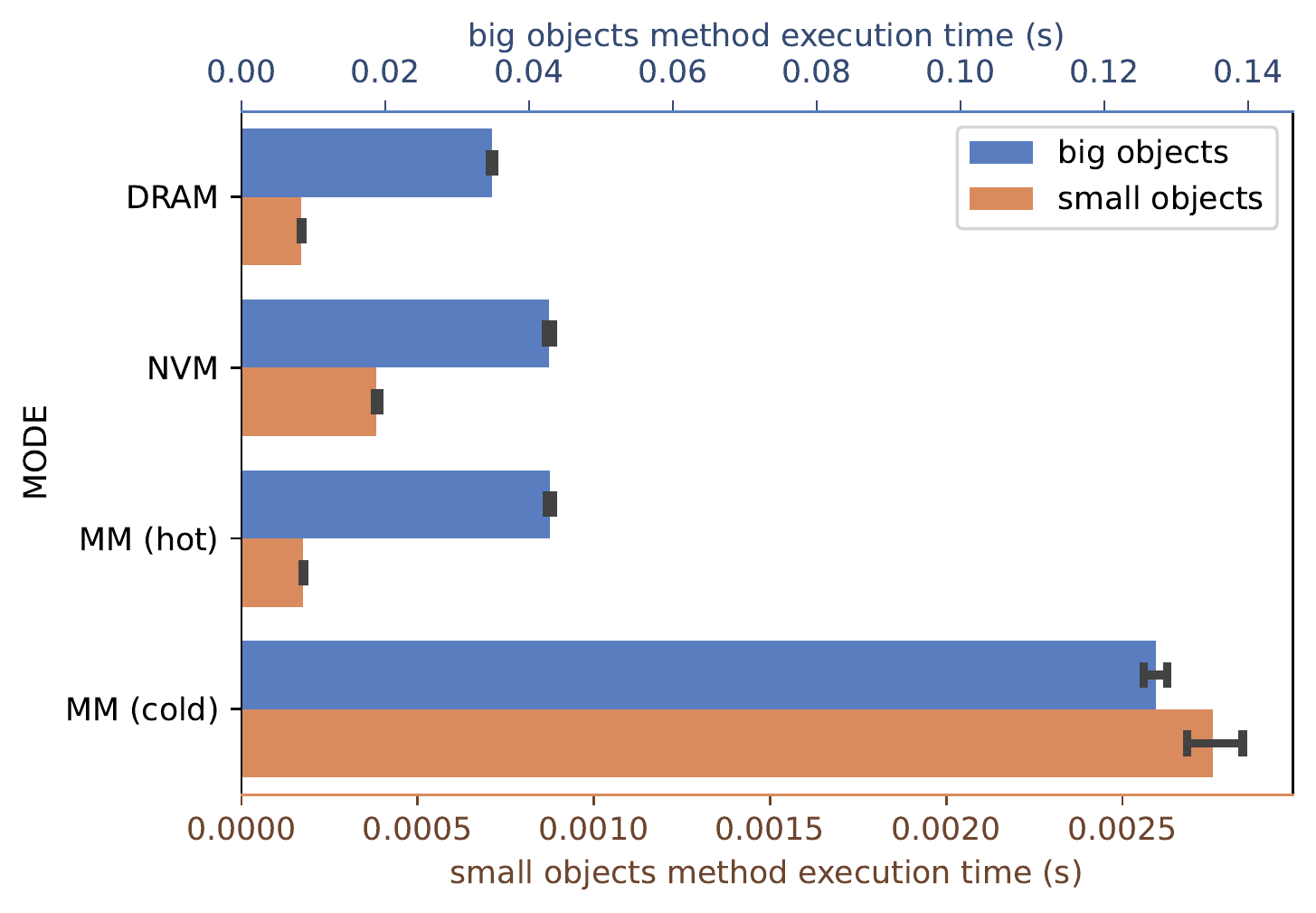}
\caption{Execution times for the matrix addition \emph{method}.\boilerplatemethod}
\label{fig:matsum-method}
\end{figure}

As done with previous applications, we include the \emph{method} execution times in Figure~\ref{fig:matsum-method}. These results illustrate the high cost of write operations onto the NVM for the following reason: the first three execution modes (\emph{DRAM}, \emph{NVM}, and \emph{MM (hot)}) are very similar in performance, but they read the input submatrices from different memory spaces (\emph{DRAM} reads them from system memory, \emph{NVM} reads them from the NVM, and \emph{MM (hot)} is expected to use DRAM as it is running \emph{hot}). However, they all have in common that they write the output object to DRAM. The last mode, \emph{MM (cold)}, shows a much worse performance due to the fact that it is writing the output object to NVM, which results in a noticeable penalty.

\paragraph{Insights}
There are two main ideas that can be drawn from this application, ideas that can be interpolated to many memory-bound applications. 
First of all, data locality (and thus, an active storage system) can bring great benefits to such kind of application thanks to the reduction of data transfers and memory copies --memory operations that are the bottleneck of memory-bound applications. 
Secondly, one can generally expect a trade-off between the memory footprint and performance for such kind of applications. Given the NVM high write cost, it is beneficial to perform all write operations onto fastest memory (in our specific scenario, that means having the output dataset in DRAM). 
Just as already discussed in the previous applications, we can observe how all active scenarios outperform the non-active one. The active object store has some additional potential which, in this case, allows for a further performance improvement by leveraging both memory tiers, which sidesteps the bottleneck of write bandwidth.

\subsection{Matrix Multiplication}
\label{evaluation:matmul}

\begin{figure}
\includegraphics[width=\columnwidth]{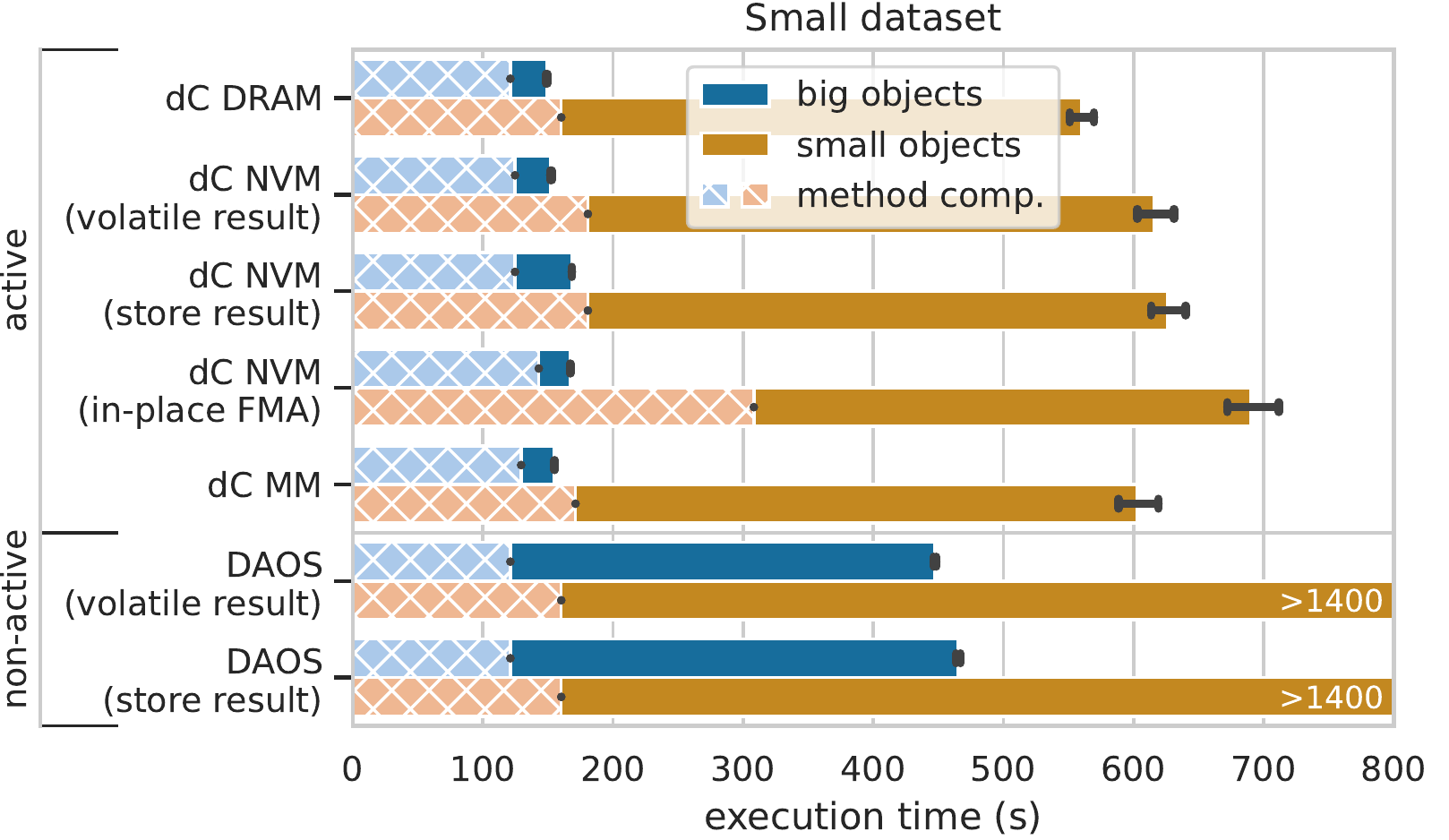}
\caption{Execution times of the matrix multiplication application for a small dataset ($n = 42000$).\boilerplateappsmall{} The result may be held in DRAM (\emph{volatile result} executions) or stored in the object store (\emph{store result} executions). The \emph{store result} execution corresponds to an execution which evaluates the output object in DRAM and, after the \emph{active method} invocation has finished, transfers it to the object store. The other execution, \emph{in-place FMA}, performs all the operation in-place in the NVM, without having any input or output structure in DRAM.}
\label{fig:matmul-small}
\end{figure}

\begin{figure}
\includegraphics[width=\columnwidth]{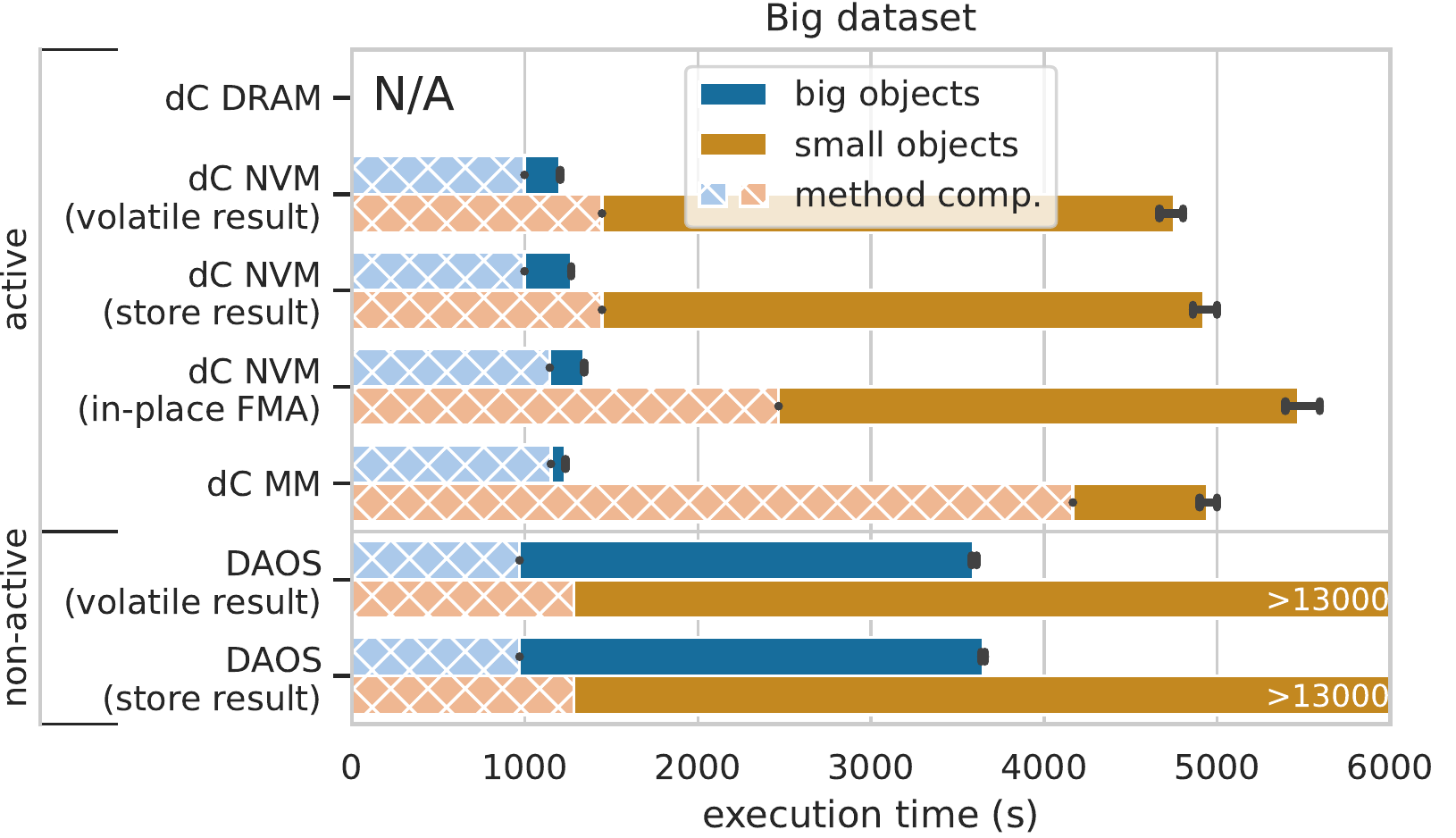}
\caption{Execution times of the matrix multiplication application for a big dataset ($n = 84000$).\boilerplateappbig{} The result may be held in DRAM (\emph{volatile result} executions) or stored in the object store (\emph{store result} executions). The \emph{store result} execution corresponds to an execution which evaluates the output object in DRAM and, after the \emph{active method} invocation has finished, transfers it to the object store. The other execution, \emph{in-place FMA}, performs all the operation in-place in the NVM, without having any input or output structure in DRAM.}
\label{fig:matmul-big}
\end{figure}

The different execution times can be seen in Figure~\ref{fig:matmul-small} (small dataset) and Figure~\ref{fig:matmul-big} (big dataset).
The data structures of Matrix Multiplication (input dataset, output dataset, and objects) are identical to the previous application, Matrix Addition; however, the memory access pattern and the computation requirements are completely different.

We can see a consistent good behavior of the active object store, specially when using big objects where active executions are 3 times faster than the non-active ones. The object reuse is something that amplifies the costs of the non-active approach, an aspect that we had already seen in the \emph{k}-means application (Section~\ref{evaluation:kmeans}).

There is a huge difference between executions with big objects and small objects: active executions are almost 5 times slower when using small objects than the ones using big objects. This is due to an increase on the \emph{active method} invocation count and general data reuse: that factor is $6\times$ for big objects and $42\times$ for small objects as shown in Table~\ref{tbl:app-comparison}; those factors are a consequence of the different object size.

In the previous Matrix Addition application we had observed a big impact when having the output dataset volatile or persisted in the NVM. This difference is softened and it becomes almost inappreciable. In this case, the long computations required by the Matrix Multiplication (shown by the high computation to data ratio in Table~\ref{tbl:app-comparison}) decreases the gap between the different memory configurations --they all are within 15\% between one another for the small objects, and only 10\% for big objects. This remains true even while doing a write-heavy multiply\hyp{}and\hyp{}add operation in-place in the NVM (labeled \emph{in-place FMA}); this configuration is more taxing regarding write operations on the NVM but it yields a small overhead in relative terms.

We have already discussed the programming and development aspect of the volatile vs store placement of the result (see Matrix Addition evaluation \ref{evaluation:matsum}), and the same considerations apply for the in-place FMA configuration.

The fastest memory configuration is to store the output Matrix in DRAM (\emph{volatile result}). However, the decreased gap between executions implies that we can perform a huge Matrix Multiplication with virtually zero DRAM footprint with a bearable performance degradation, as shown by the \emph{in-place FMA} execution times.

\begin{figure}
\centering
\includegraphics[width=\columnwidth]{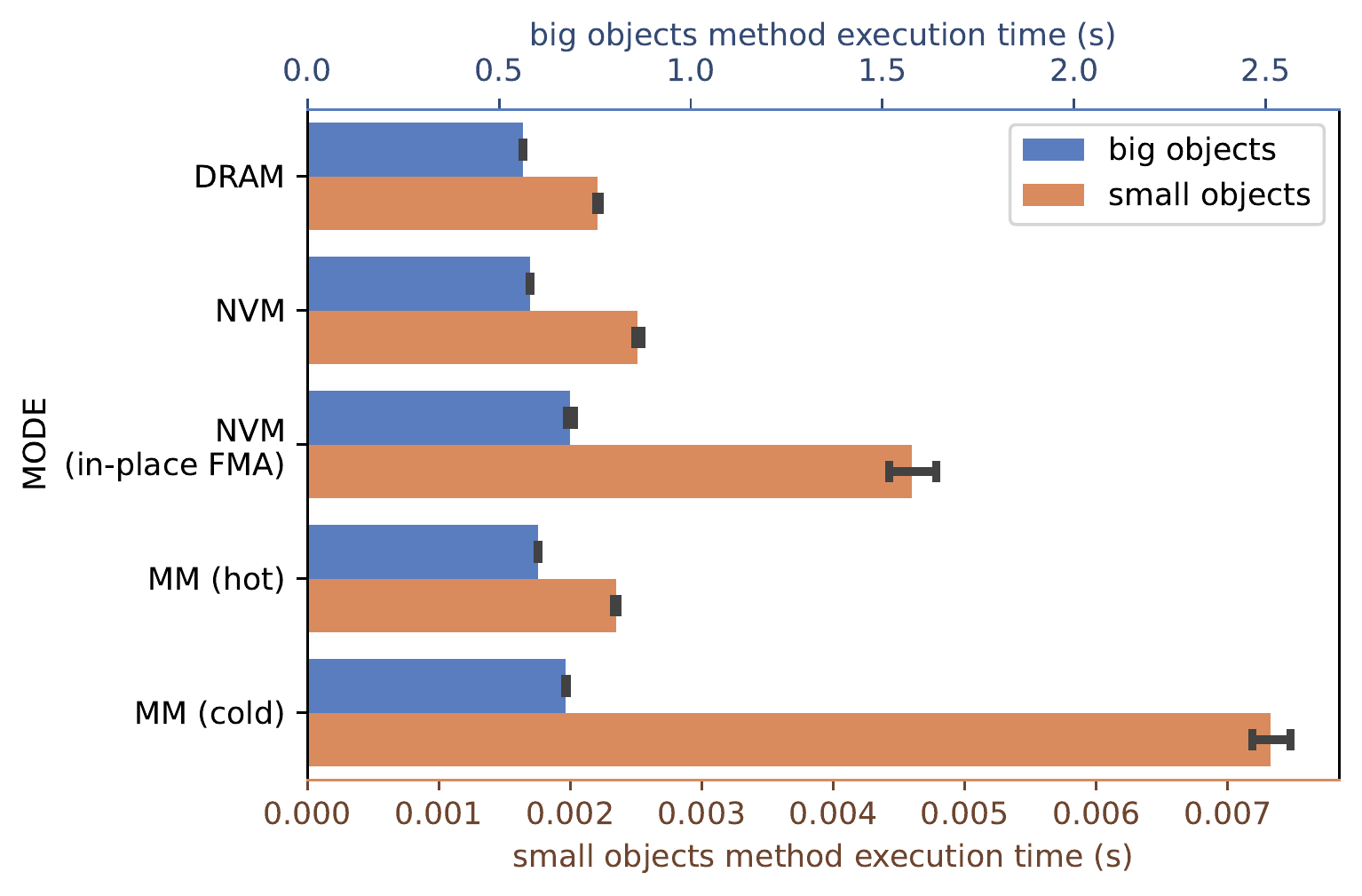}
\caption{Execution times for the matrix multiplication \emph{method}.\boilerplatemethod{} The \emph{method} may perform the fused\hyp{}multiply\hyp{}add operation in DRAM or in-place on the NVM; this last execution corresponds to the \emph{in-place FMA} one in the figure.}
\label{fig:matmul-method}
\end{figure}

As in previous applications, we include the \emph{method} execution time analysis in Figure~\ref{fig:matmul-method}. The \emph{method} execution when done with big objects is almost the same (within 10\%) across all configurations, something that matches the behavior of the application and is explained due to the high computation to data ratio. 
For small objects, where the absolute execution times are lower, the impact of the memory configuration is much more visible; in both object sizes \emph{NVM (in-place FMA)} and \emph{MM (cold)} are the slower configurations, but the relative difference is much more apparent when using small objects which are between $2\times$ and $3\times$ slower.

\paragraph{Insights}
This application has shown how, for scenarios with high computation requirements and significant data reuse, the data locality plays a main role in the overall performance while the exact memory configuration is secondary. Even more important than the available memory and its configuration is the object size, which will impact the quantity of \emph{active method} invocations and it can have a dramatic impact on overall performance.

\subsection{Known limitations}

\added{
In this section we have evaluated the potential of active object stores by evaluating performance results, in a controlled environment, with four different applications. However, it is relevant to remember certain limitations that such designs have.
}

\added{
First of all we have observed widely different improvements depending on the application. This is somewhat expected: performance results during NVM in-place execution will depend on the data access pattern. However, quantifying \emph{a priori} this gain can be challenging, specially for complex real\hyp{}life applications.
}

\added{
If we look into the hardware requirements of the storage system we can observe that the software stack needs both storage and computing capabilities. This may pose a problem, as a lot of traditional infrastructures segregate nodes into storage nodes (with almost no spare computing capabilities) and computing nodes (which may not have enough storage). This hinders the ease of adoption of active storage systems.
}

\added{
Having intertwined storage and computing resources results in more complex management and more fuzzy price~/~performance analysis: scaling up a \emph{hyper\hyp{}converged} infrastructure is costlier than scaling up either compute nodes or storage nodes. This may become a deal\hyp{}breaker for an infrastructure migration plan.
}

\added{
If the active object store is not deployed in an ephemeral configuration --i.e., instead of being deployed in the same nodes as the application is deployed in a set of nodes and shared between different applications concurrently-- then interference problems are expected to arise and become a problem. This is specially true in time\hyp{}critical scenarios, in the presence of SLA, or in similar settings.
}

\section{Conclusions}

\label{sec:conclusions}

In this article we have proposed an active object store design (based on dataClay) and evaluated the potential and improvements that such storage system design brings. The software stack has proven to be versatile, and it presents very good performance results. We have performed the evaluation with different applications which represent a vast set of applications in HPC, data analytics, and machine learning domains.

As the evaluation has shown, the benefits obtained by this design will vary depending on the application. However, in all  scenarios, we have observed that the active features provide substantial gains. We have experiments that show 10\% improvement for an application with no data reuse and high computation data ratio (represented by the Histogram) and up to an order of magnitude for a memory-bound application under the most appropriate memory configuration (Matrix Addition with the result placed in DRAM).

On top of the aforementioned benefits, obtained through the \emph{active} capabilities of the object store, we have also shown different memory configurations and the impact that an NVM device can have. The byte\hyp{}addressable nature of this memory tier enables in-place execution; we have demonstrated how the active object store is able to execute applications with datasets bigger than system memory at near-DRAM speed.

After the evaluation we can conclude that there are two main data locality improvements that are considered: on the one hand, the \emph{active} capabilities avoid any transfers between the storage system and the application --that is regardless of the exact memory configuration or the presence of NVM devices. On the other hand, the byte\hyp{}addressable memory avoids data copies between memory tiers (DRAM and NVM) within the storage system itself. This last aspect shows a great potential but can also backfire, as it is a slower memory and for instance the \emph{k}-means has shown how an explicit memory copy can improve the performance by a $2\times$ factor (\emph{pre-copy} strategy).

As previously stated, memory-bound applications have great performance improvements but there are also gains for compute-bound applications as well. An example of such kind of application is showcased in the matrix multiplication application. Since the data locality overall impact is reduced due to the higher computation cost, compute-bound applications show lower sensitivity to memory speeds.
We have seen how those kind of applications can be run almost entirely from within the NVM (i.e., huge datasets, with minimal memory footprint) while performance degradation is as little as 10\% in our experiments --and still more than $2\times$ faster than non-active executions.

The evaluation of the storage system is done with an ephemeral configuration of dataClay, a setup that is available and ready to be deployed. We believe that both ephemeral storage systems and NVM can be game\hyp{}changers agents on computation clusters (at the software and at the hardware level, respectively). In this article we have shown some great performance results, which showcase not only the potential of \emph{active} object stores but also the opportunities that these designs bring when combined with other technologies such as NVM.

\section{Acknowledgements}

This work has been partially supported by the Spanish Government
(PID2019-107255GB), by Generalitat de Catalunya (contract
2014-SGR-1051), and by the European Commission through the Horizon 2020
Research and Innovation program and the EuroHPC JU under contract 955558
(eFlows4HPC project). Anna Queralt is a Serra H\'unter Fellow. This work
has been partially supported by Intel\textregistered{} under the BSC-Intel collaboration on 3D XPoint\texttrademark{} Technology.

\bibliographystyle{ACM-Reference-Format}
\bibliography{references.bib}

\end{document}